\documentclass[twocolumn]{aastex61}
\usepackage{color}

\newcommand{\Msun}{\mbox{$M_{\odot}$}}

\shorttitle{The Elusive Majority of Young Moving Groups}
\shortauthors{Bowler et al.}

\begin{document}

\title{The Elusive Majority of Young Moving Groups. I. \\ Young Binaries and Lithium-Rich Stars in the Solar Neighborhood}

\correspondingauthor{Brendan P. Bowler}
\email{bpbowler@astro.as.utexas.edu}

\author[0000-0003-2649-2288]{Brendan P. Bowler}
\altaffiliation{Visiting astronomer, Kitt Peak National Observatory, National Optical Astronomy Observatory, which is operated by the Association of Universities for Research in Astronomy (AURA) under a cooperative agreement with the National Science Foundation. }
\altaffiliation{Visiting astronomer, Cerro Tololo Inter-American Observatory, National Optical Astronomy Observatory, which is operated by the Association of Universities for Research in Astronomy (AURA) under a cooperative agreement with the National Science Foundation.}
\affiliation{Department of Astronomy, The University of Texas at Austin, Austin, TX 78712, USA}

\author{Sasha Hinkley}
\affiliation{University of Exeter, Physics and Astronomy, EX4 4QL Exeter, UK}

\author{Carl Ziegler}
\affiliation{Dunlap Institute for Astronomy and Astrophysics, University of Toronto, Toronto, Ontario M5S 3H4, Canada}

\author{Christoph Baranec}
\affiliation{Institute for Astronomy, University of Hawai`i at M\={a}noa, 640 N. A'oh\={o}k\={u} Pl., Hilo, HI 96720, USA}

\author{John E. Gizis}
\affiliation{Department of Physics and Astronomy, University of Delaware, Newark, DE 19716, USA}

\author{Nicholas M. Law}
\affiliation{Department of Physics and Astronomy, University of North Carolina at Chapel Hill, Chapel Hill, NC 27599-3255, USA}

\author{Michael C. Liu}
\affiliation{Institute for Astronomy, University of Hawai`i at M\={a}noa, 2680 Woodlawn Drive, Honolulu, HI 96822, USA}

\author{Viyang S. Shah}
\affiliation{Department of Astronomy, The University of Texas at Austin, Austin, TX 78712, USA}

\author{Evgenya L. Shkolnik}
\affiliation{School of Earth and Space Exploration, Arizona State University, Tempe, AZ 85281, USA}

\author{Basmah Riaz}
\affiliation{Universit\"{a}ts-Sternwarte M\"{u}nchen, Ludwig Maximilians Universitat, Scheinerstra{\ss}e 1, D-81679 M\"{u}nchen, Germany}

\author{Reed Riddle}
\affiliation{California Institute of Technology, 1200 E. California Blvd., Pasadena, CA 91125, USA}

\begin{abstract}

Young stars in the solar neighborhood serve as nearby probes of stellar evolution 
and represent promising targets to directly image self-luminous giant planets.
We have carried out an all-sky search for late-type ($\approx$K7--M5) 
stars within 100 pc selected primarily on the basis of activity indicators from $GALEX$ and $ROSAT$.
Approximately two thousand active and potentially young stars are identified, over 600 of which 
we have followed up with low-resolution optical spectroscopy and over 1000 with diffraction-limited imaging using 
Robo-AO at the Palomar 1.5-m telescope.  
Strong lithium is present in 58 stars, implying ages spanning $\approx$10--200 Myr.  
Most of these lithium-rich stars are new or previously known members of young moving groups including
TWA, $\beta$ Pic, Tuc-Hor, Carina, Columba, Argus, AB Dor, Upper Centaurus Lupus, and Lower Centaurus Crux; the rest  
appear to be young low-mass stars without connections to established kinematic groups.  
Over 200 close binaries are identified down to 0$\farcs$2 --- the vast majority of which are new --- and 
will be valuable for dynamical mass measurements of young stars with continued orbit monitoring in the future.

\end{abstract}

\keywords{binaries: visual---stars: low-mass---stars: pre-main sequence}

\section{Introduction}{\label{sec:intro}}

Since the initial recognition of young moving groups (YMGs) about two decades ago 
(e.g., \citealt{Kastner:1997fk}; \citealt{Zuckerman:2000gf}; \citealt{Torres:2000kn}), 
these nearby associations of intermediate-age ($\approx$10--200 Myr) stars
have been the subject of increasing interest in the stellar,
substellar, and exoplanet communities (e.g., \citealt{Torres:2008vq}; \citealt{Mamajek:2016ik}; \citealt{Bowler:2016jk}).
These loose, relatively sparse ($N$$\sim$50--300), kinematically comoving groups 
in the solar neighborhood ($\lesssim$100 pc) provide a link between 
the youngest T Tauri stars and the older population of field stars.

Because of their proximity and youth, YMGs have become a rich resource to study
a broad range of topics:
the evolution of stellar dynamos and activity (e.g., \citealt{Shkolnik:2014jl}; \citealt{Ansdell:2015bc});
dynamical masses of intermediate-age stars 
(e.g., \citealt{Close:2005kx}; \citealt{Montet:2015ky}; \citealt{Nielsen:2016ct}; \citealt{Janson:2018fc});
the structure and evolution of debris disks (e.g., \citealt{Wyatt:2014up});
young brown dwarfs and free-floating planetary-mass objects
(\citealt{Liu:2013gya}; \citealt{Allers:2013hk}; \citealt{Gagne:2014gp}; \citealt{Aller:2016kg}; \citealt{Liu:2016co}; \citealt{Faherty:2016fx}); 
multiplicity at young ages (\citealt{Shan:2017bn}; \citealt{Janson:2017hl}; \citealt{Best:2017bra});
and the initial mass function of sparse clusters (\citealt{Gagne:2017gy}).
Members of YMGs have also become favored targets for 
direct imaging searches for exoplanets 
(e.g., \citealt{Biller:2013fu}; \citealt{Brandt:2014hc}; \citealt{Bowler:2015ja}; \citealt{Chauvin:2015jy}) and,
as a result, many of the known directly imaged planets and planetary-mass companions orbit 
members of these associations  
(e.g., 2M1207--3932~b, \citealt{Chauvin:2004cy}; HR 8799~bcde, \citealt{Marois:2008ei}; 
$\beta$~Pic~b, \citealt{Lagrange:2010fsa}, 51~Eri~b, \citealt{Macintosh:2015ewa}; 
GU Psc~b, \citealt{Naud:2014jx}; 2M2236+4751~b, \citealt{Bowler:2017hq}).
However, the relatively limited number of \emph{bona fide} members of young moving 
groups---a few hundred confirmed using fully-constrained space motions together with other independent 
youth indicators---has gradually 
become a barrier to measuring more precise occurrence rates with direct imaging and searching for
correlations with stellar host mass (\citealt{Bowler:2018dq}).

Despite numerous dedicated searches to identify nearby young stars, the current census of 
stellar and substellar members of YMGs is vastly incomplete.
Assuming a standard initial mass function, 
\citet{Kraus:2014ur}, \citet{Gagne:2017gy}, and \citet{Shkolnik:2017ex}
find that tens to hundreds of low-mass stars and brown dwarfs are probably missing from  membership lists 
of Tuc-Hor, TWA, and $\beta$ Pic.  The same is likely to be true of other YMGs owing to early, biased searches 
for bright members using \emph{Hipparcos} parallaxes and proper motions.  
This has prompted a number of programs to find new low-mass members spanning
the stellar and substellar mass regimes
(\citealt{Gizis:2002je}; \citealt{Lepine:2009ey}; \citealt{Shkolnik:2009dx}; \citealt{Schlieder:2010gka}; \citealt{Schlieder:2012gj}; 
\citealt{Malo:2013gn}; \citealt{Malo:2014dk}; \citealt{Gagne:2014gp}; \citealt{Kraus:2014ur}; \citealt{Riedel:2014ce}; \citealt{Binks:2015fc}; 
\citealt{Aller:2016kg}; \citealt{Riedel:2017el}; \citealt{Shkolnik:2017ex}).
In spite of these innovative efforts, hundreds of low-mass members likely await discovery.

Motivated by the need for additional targets for high-contrast imaging, we have carried out a broad search for 
low-mass stars in young moving groups.  
The goals of this program are highly focused: to identify new, single, relatively bright ($R$ $\lesssim$ 15~mag) YMG members 
with large proper motions.  This facilitates the rapid discrimination of background stars from bound companions 
for follow-up high-contrast imaging observations.
Our strategy is to initially use X-ray and UV activity together with  color and proper motion cuts 
to locate candidate young early-M dwarfs.
Having begun this study prior to $Gaia$ data releases, our approach to selecting targets for follow-up observations 
has relied only on proper motions and sky positions without the advantage of having parallaxes.

This study focuses on the characterization of potential young stars and moving group members
based on 
low-resolution optical spectroscopy 
together with adaptive optics imaging with Robo-AO at the Palomar 60-inch (1.5-m) telescope.
In a separate paper we will present radial velocities from new high-resolution spectroscopy 
of several hundred 
potential moving group members 
as part of a follow-up kinematic analysis.
Section 2 summarizes the activity, color, and proper motion cuts used to define our starting sample.
Our observations and analysis are described in Sections 3 and 4.
Moving group candidates are discussed in Section 5 and our conclusions are summarized in Section 6.

\section{Sample Selection}{\label{sec:sample}}

Our starting sample draws from two large catalogs of low-mass stars.  
The \citet{Frith:2013dt} list of bright M dwarfs ($K$$<$9~mag)
consists of stars between K7--M4 selected from the PPMXL catalog (\citealt{Roeser:2010cr}).
The authors apply a series of optical and NIR color cuts to isolate late spectral types, and 
reduced proper motions are used to distinguish dwarfs from bright, distant giants.  Frith et al. require a S/N of
at least 5 for proper motions and remove regions surrounding the galactic plane ($|b|$ $<$ 15$^{\circ}$) 
susceptible to source confusion.  Finally, they combine their list with the \citet{Lepine:2011gl}
catalog of bright M dwarfs to produce a final catalog of 8479 late-K to mid-M dwarfs.


\begin{figure}
  \vskip -.2 in
  \hskip -.4 in
  \resizebox{4.0in}{!}{\includegraphics{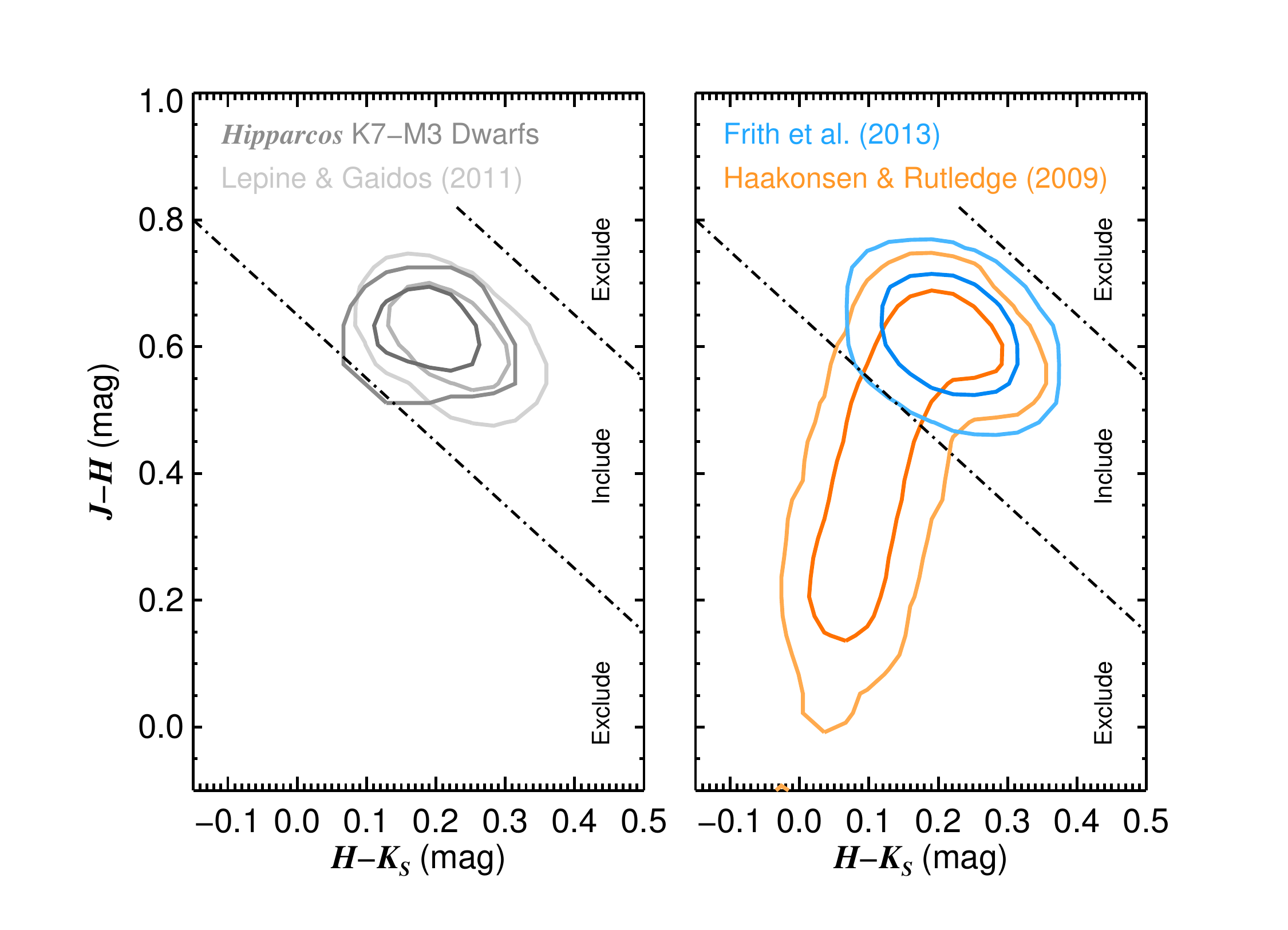}}
  \caption{Near-infrared color cuts applied to the F13 and H09 catalogs to isolate 
  late-K to mid-M dwarfs (dot-dashed lines).
  \emph{Left}: Comparison samples of early-M dwarfs from \citet[light gray]{Lepine:2011gl} and 
  the XHIP extended compilation of \emph{Hipparcos} 
  stars from \citet[dark gray]{Anderson:2012cu}. 
  \emph{Right}: The F13 catalog (blue) is already selected for M dwarfs, but earlier spectral types are
  excluded from the H09 sample (orange) with these color cuts. 
  Contours encompass 68\% and 95\% of objects with near-infrared photometric uncertainties $<$0.1~mag.
   \label{fig:nirccd} } 
\end{figure}


\begin{figure}
  \vskip -.2 in
  \hskip -.4 in
  \resizebox{4.0in}{!}{\includegraphics{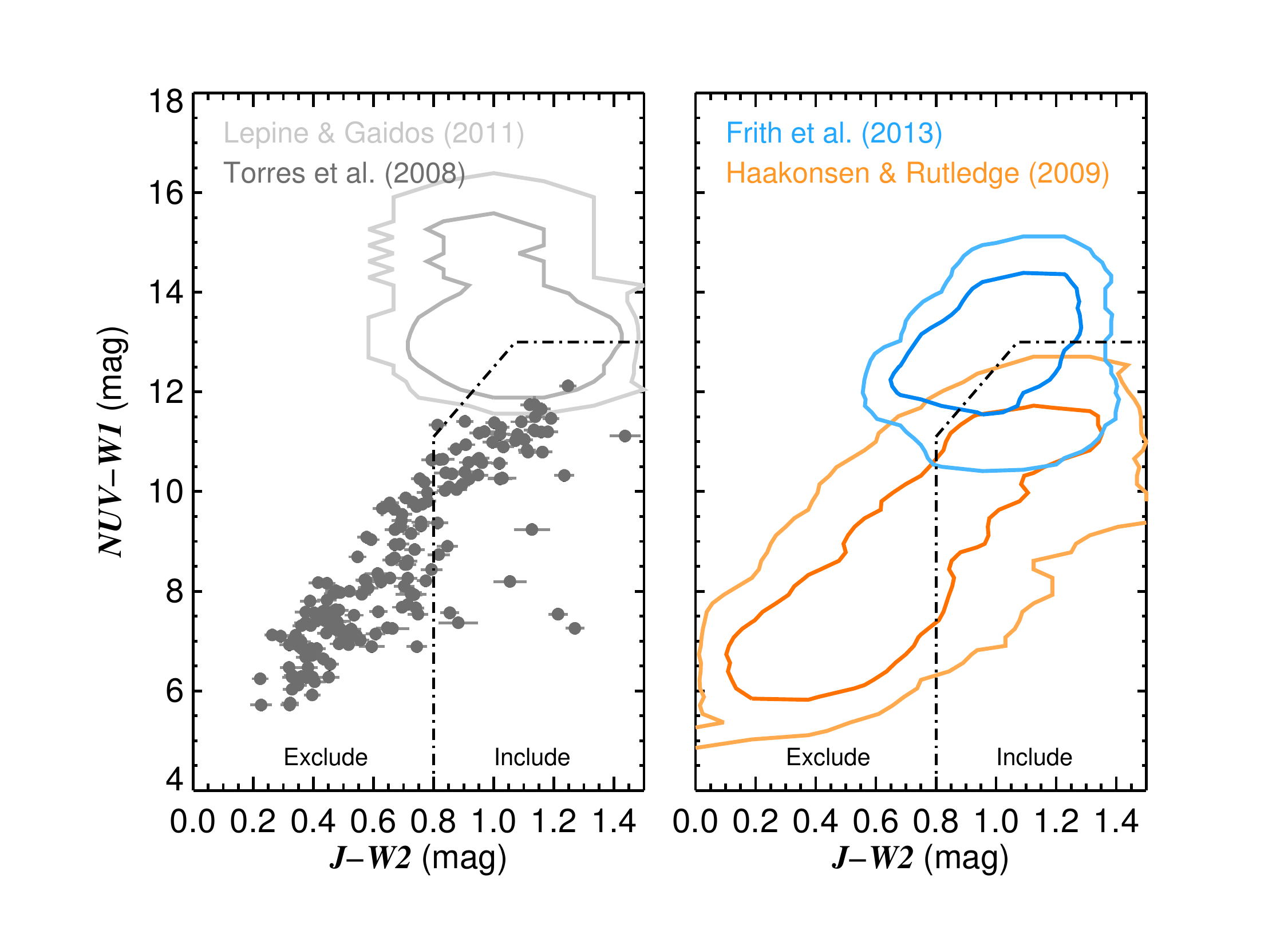}}
  \caption{Activity cuts using $NUV$--$W1$ and $J$--$W2$ photometry (dot-dashed lines).  
  \emph{Left}: Comparison sample of field M dwarfs from \citet{Lepine:2011gl} together with the compilation 
  of known YMG members from \citet{Torres:2008vq} spanning 10--150~Myr.  Most YMG members trace
  out a saturated locus of NUV emission compared to the field population at a given $J$--$W2$ color, which is a proxy
  for spectral type.  Late-K and M dwarfs have $J$--$W2$ colors $\gtrsim$0.8 mag.
  \emph{Right}: Our color cuts applied to the F13 and H09 samples.  Most of the F13 M dwarfs are relatively
  inactive, whereas the H09 stars are pre-selected to also exhibit X-ray emission and are therefore also UV bright.
     \label{fig:activitycut} } 
\end{figure}

We also utilize the \citet{Haakonsen:2009iq} list of $ROSAT$ All-Sky Survey Bright Source Catalog (\citealt{Voges:1999ws}) detections 
cross-matched with the 2MASS Point Source Catalog (\citealt{Cutri:2003tp}; \citealt{Skrutskie:2006hl}).   The authors provide probabilities that each
X-ray source is uniquely associated with a near-infrared counterpart.  Altogether 18,497 ROSAT detections have
non-zero probabilities of being associated with a 2MASS sources.  For this study we select 6084 
targets with $>$90\% association probabilities as a supplementary catalog to search for young active M dwarfs.

Both samples are then cross-matched against all-sky photometric and proper motion surveys.
Near-infrared $J$-, $H$-, and $K_S$-band photometry is extracted from the 
Two Micron All Sky Survey (2MASS; \citealt{Skrutskie:2006hl}) with a search
radius ($R_S$) of 5$''$; $r'$-band photometry is from Carlsberg Meridian Catalogue 14 (\citealt{Evans:2002gy}; $R_S$=5$''$);  
$R2$ magnitudes are from USNO-B1.0 (\citealt{Monet:2003bw}; $R_S$=5$''$);
$NUV$ and $FUV$ photometry is from the latest \emph{Galaxy Evolution Explorer} ($GALEX$) 
General Release (GR6/GR7; \citealt{Martin:2005ko}; \citealt{Morrissey:2007ch}; $R_S$=10$''$); 
\emph{W1}, \emph{W2}, \emph{W3}, and \emph{W4} photometry from \emph{Wide-field Infrared Survey Explorer} 
($WISE$; \citealt{Wright:2010in}; $R_S$=10$''$); 
X-ray count rates and hardness ratios are from the $ROSAT$ All-Sky Survey Bright Source Catalog (\citealt{Voges:1999ws}) or,
if not detected there, then the $ROSAT$ All-Sky Faint Source Catalog (\citealt{Voges:2000wn}; $R_S$=30$''$);
and $V$-band magnitudes and proper motions are from the USNO CCD Astrograph Catalog 4 (\citealt{Zacharias:2013cf}; $R_S$=5$''$).
If there are multiple $GALEX$ detections for the same search position at different epochs then we adopt the weighted mean and 
uncertainty of these measurements.  

We apply a series of color, activity, proper motion, and photometric distance cuts to both catalogs that are specifically designed to identify nearby 
young M dwarfs for follow-up planet searches with direct imaging.  These criteria are primarily intended 
for the \citet{Haakonsen:2009iq} catalog (hereinafter HR09),
which has a diverse mix of non-stellar ``contaminants''  (active galactic nuclei, cataclysmic variables, galaxy clusters, etc.).  
On the other hand, the \citet{Frith:2013dt} catalog (hereinafter F13) is well-vetted for M dwarfs, but these are overwhelmingly 
expected to be old inactive field stars.  
Below we list the additional filters we have applied to both samples:

\begin{itemize}
\item \textbf{Optical brightness cut.}  Stars with $r'$ $>$ 15 mag are excluded.  This corresponds to the approximate faintness limit for 
natural guide star AO instruments like Keck/NIRC2, ensuring an optically-bright sample for the possibility of 
follow-up high-contrast imaging.  If no $r'$ measurement is 
listed in CMC14 then we adopt the $R2$ magnitude from USNO-B1.0 and apply the same brightness cut.

\item \textbf{Photometric distance cut.}  $V$-band photometric distance estimates are computed using the $M_V$ versus $V$--$K_S$ band
polynomial fit to Pleiades stars in \citet{Bowler:2013ek}.  
Most known moving group are located within about 100~pc, so we further restrict our search catalog to photometric distances $<$ 100 pc.
Photometric distances will underestimate the true distances for 
binaries and young stars still descending along the Hayashi track, but this cut excludes most of the distant M dwarfs from the sample.

\item \textbf{Near-infrared color cuts.}  
A series of near-infrared color cuts are imposed to further isolate late-K and early-M dwarfs.
Only stars with $J$-band, $H$-band, and $K_S$-band photometric uncertainties below 0.1~mag are considered.
\emph{Hipparcos} K7V--M3V stars and the \citet{Lepine:2011gl} sample of 
bright M dwarfs are used to establish typical near-infrared colors of M dwarfs (Figure~\ref{fig:nirccd}).
Based on this locus, we impose the following color cuts:
\begin{equation}
J - H  > -(H-K_S) + 0.65 \mathrm{\ mag}
\end{equation}
\begin{equation}
J-H < -(H-K_S) + 1.05 \mathrm{\ mag}
\end{equation}

These cuts are depicted in Figure~\ref{fig:nirccd} for two control samples from \emph{Hipparcos}
and \citet{Lepine:2011gl}, in addition to the F13 and H09 catalogs we consider in this work.
M dwarfs have already been color-selected for the F13 catalog, so this cut predominantly affects 
the H09 catalog.

\item \textbf{UV activity cut.}  
Stars with active chromospheres are readily distinguished from their inactive 
counterparts using $GALEX$ photometry.
Following \citet{Rodriguez:2013fv}, we use the $J$--$W2$ versus $NUV$--$W1$ diagram to identify
active stars (Figure~\ref{fig:activitycut}):
\begin{equation}
NUV - W1  < 7.0 (J-W2) + 5.5 \mathrm{\ mag}
\end{equation}
\begin{equation}
NUV-W1 < 13 \mathrm{\ mag}
\end{equation}
Based on the spectral type-color relation from \citet{Rodriguez:2013fv}, we also 
require that $J$--$W2$ $>$ 0.8~mag to isolate late-type ($\ge$K5) stars
(Figure~\ref{fig:activitycut}). 
Note that this cut does not remove white dwarf-M dwarf binaries, which can share
similar UV-to-infrared colors as young, active M dwarfs (\citealt{Silvestri:2007bx}; \citealt{Shkolnik:2011in})

\item \textbf{Reduced proper motion cut.}  
Reduced proper motions provide a convenient way to separate fast-moving dwarfs from 
kinematically slow but luminous giants.  Following F13, we require $H_K$ $>$ 6.0,
where the reduced proper motion is
$H_K$ = $K$ + 5 log($\sqrt(\mu_{\alpha}\cos(\delta)^2 + \mu_{\delta}^2))$ + 5;
here $\mu_{\alpha}$$\cos(\delta)$ and $\mu_{\delta}$ are the star's proper motion 
in arcseconds per year.  Finally, we also require the total proper motion to be greater 
than 25 mas yr$^{-1}$ to ensure that candidate planets identified in AO imaging
can be distinguished from background stars on short ($\sim$1~year) timescales.

\end{itemize}

Cross-matching the resulting filtered F13 and H09 samples 
yields 2060 unique targets which we use as the starting point for our young moving group kinematic selection.

\section{Observations}{\label{sec:obs}}

To better characterize our starting sample of 2060 activity-selected late-K and early-M dwarfs,
we carried out a follow-up observational program to obtain low-resolution optical spectra of these targets using 
instruments in the northern and southern hemispheres, together with AO imaging with Robo-AO at 
the Palomar 60$''$ (1.5-m) telescope in the north.
Altogether we acquired 762 optical spectra of 632 stars, plus an additional four nearby stars sharing common
proper motions with targets in our sample.
We also obtained 1523 AO images of 1011 stars to uncover and characterize close binaries.
The broader goals of this program are to identify single young stars for high-contrast imaging, so 
known binaries from recent high-resolution campaigns (e.g., \citealt{Janson:2012dc}; \citealt{Janson:2014fx})
are deprioritized, leading to an intentionally biased sample which we note is not easily amenable to multiplicity statistics.
Details about the instrument setups and data reduction are discussed below.

\begin{deluxetable*}{lccccccccccc}
\renewcommand\arraystretch{0.9}
\tabletypesize{\footnotesize}
\setlength{ \tabcolsep }{.1cm}
\tablewidth{0pt}
\tablecolumns{12}
\tablecaption{Spectroscopic Observations\label{tab:specobs}}
\tablehead{
       \colhead{2MASS} & \colhead{Date}  & \colhead{Telescope/}  & \colhead{}           & \colhead{Res.}   & \colhead{Exp.} & \colhead{H$\alpha$ EW}           & \colhead{Li EW}                 & \colhead{Na EW}                   & \colhead{TiO5} & \colhead{Hammer} & \colhead{Vis.} \\
       \colhead{Name} & \colhead{(UT)}   & \colhead{Instrument}  & \colhead{Grating}    & \colhead{Power}  & \colhead{(s)}      & \colhead{(\AA)\tablenotemark{a}} & \colhead{(\AA)\tablenotemark{a}} & \colhead{(\AA)\tablenotemark{a}} & \colhead{Index} & \colhead{SpT\tablenotemark{b}} & \colhead{SpT\tablenotemark{b}} 
  }
  \startdata
J00022714--4601439    & 2014--06--27 &      SOAR/Goodman   &    SYZY400 &  1800 &   300   &     --1.0  &  $\cdots$ &    2.2   &  0.65 &         M1  &       M2  \\ 
J00104302--2039067    & 2013--12--06 &      SOAR/Goodman   &    SYZY400 &  1800 &   120   &     --3.5  &  $\cdots$ &    3.1   &  0.48 &         M3  &       M3  \\ 
J00104302--2039067    & 2013--12--06 &      SOAR/Goodman   &   RALC1200 &  5900 &   240   &     --3.2  &  $\cdots$ & $\cdots$ &  0.49 &         M3  &       M3  \\ 
J00114643--1139553    & 2013--12--06 &      SOAR/Goodman   &    SYZY400 &  1800 &   150   &      0.3  &  $\cdots$ &    3.6   &  0.75 &         M0  &       M0  \\ 
J00120761--1550327    & 2013--12--31 &      Mayall/RC-Spec &     BL420  &  2600 &    30   &     --3.2  &  $\cdots$ &    1.1   &  0.93 &        K:  &     G/K:  \\ 
J00141709--6139237    & 2013--12--05 &      SOAR/Goodman   &    SYZY400 &  1800 &   300   &     --1.7  &  $\cdots$ &    3.1   &  0.52 &         M2  &       M3  \\ 
J00141709--6139237    & 2013--12--05 &      SOAR/Goodman   &   RALC1200 &  5900 &   500   &     --1.8  &  $\cdots$ & $\cdots$ &  0.55 &         M2  &       M2  \\ 
J00144767--6003477    & 2014--06--25 &      SOAR/Goodman   &    SYZY400 &  1800 &   300   &     --5.1  &  $\cdots$ &    3.4   &  0.39 &         M4  &       M4  \\ 
J00151561+0247373    & 2014--06--26 &      SOAR/Goodman   &    SYZY400 &  1800 &   240   &      0.4  &  $\cdots$ &    1.3   &  0.88 &         K5  &       K7  \\ 
J00153670--2946003    & 2013--12--05 &      SOAR/Goodman   &    SYZY400 &  1800 &   300   &     --9.2  &  $\cdots$ &    3.5   &  0.34 &         M4  &       M5  \\ 
\multicolumn{12}{c}{$\cdots$} \\
\enddata
\tablenotetext{a}{Negative values indicate emission.  Uncertainties are estimated to be 10\% of the quoted values.}
\tablenotetext{b}{Spectral types from \texttt{Hammer} have been shown to have a systematic offset of about one spectral subclass for cool stars.  Uncertainties are $\pm$1 subclass.  Our visual spectral types are more robust and have uncertainties of $\pm$0.5 subclasses.}
\tablenotetext{c}{Likely SB2.}
\tablenotetext{d}{Visual binary.}
\tablenotetext{e}{Common proper motion companion to a star in the parent sample.}
\tablecomments{Table 1 is published in its entirety in the machine-readable format.
      A portion is shown here for guidance regarding its form and content.}
\end{deluxetable*}

\subsection{Mayall/RC-Spec}{\label{sec:mayall}}

Observations with the RC-Spectrograph mounted on the 4-m Mayall telescope at Kitt Peak were carried out over eight
nights on UT 2013 December 29--31, UT 2014 May 21--23, and UT 2015 June 16--17.
Altogether 478 spectra were obtained for 428 stars. 
The same instrument setup was used for all observing runs: the BL420 grating in conjunction with the GG-495 filter
and 1$\farcs$5 $\times$98$''$ slit dimensions produced an average resolving power ($R \equiv \lambda/\Delta \lambda$) 
of $\approx$2600 spanning 6200--9200~\AA. 
The T2KA CCD with a gain of 1.4 e$^{-}$ ADU$^{-1}$ was used for the 2013 and 2014 runs; 
the T2KB CCD was used with a gain of 1.9 e$^{-}$ ADU$^{-1}$ during the 2015 observations.
Sky conditions were partly clear with intermittent clouds.
The slit was oriented in a fixed North-South direction throughout the nights, which means targets
observed at large hour angles suffered from wavelength-dependent slit loss from 
differential atmospheric refraction (\citealt{Filippenko:1982tl}).  
Most targets were observed near transit but the continuum slopes of some stars are affected by chromatic slit loss.
Our observations are detailed in Table~\ref{tab:specobs}.

Each image was bias subtracted, flat fielded, and corrected for bad pixels.  
Night sky lines were removed with median subtraction using 25-pix regions on either side of the science spectrum.
The spectrum was then extracted by summing the central 11-pix region in the spatial direction. 
Wavelength calibration was carried out with HeNeAr arc lamps acquired 3--5 times per night.  
About 30 prominent lines are fit with a quadratic function to derive the pixel-to-wavelength solution.
Several early-type spectrophotometric standards from \citet{Oke:1990iu}, \citet{Hamuy:1992ij},
and  \citet{Hamuy:1994bn} were observed each night to broadly correct the continuum shape 
for throughput losses from the atmosphere, optics, grating, and CCD.

\subsection{SOAR/Goodman Spectrograph}{\label{sec:soar}}

A total of 244 spectra were obtained for 168 stars with the Goodman High-Throughput Spectrograph
(\citealt{Clemens:2004je}) at the Southern Astrophysical Research (SOAR) 4.1-m telescope
located on Cerro Pach\'{o}n, Chile.
The observations spanned nine nights on three observing runs: 
UT 2013 December 4--7, UT 2014 June 25--28, and UT 2015 February 16.
Details about individual observations can be found in Table~\ref{tab:specobs}.
Our strategy was to first observe with the 400 l mm$^{-1}$ grating (``SYZY400'') in the M2 setup with the 0$\farcs$46 slit,
which produces an average resolving power of $\approx$1800 spanning 5000--9000~\AA.
For a subset of targets --- usually those showing strong H$\alpha$ emission or hints of Li absorption --- we 
also obtained a spectrum with the 1200 l mm$^{-1}$ grating (``RALC1200'') in the M5 setup with the 0$\farcs$46 slit,
which produces an average resolving power of $\approx$5900 spanning 6250--7500~\AA.
The slit was rotated to parallactic angle for each target on all nights except UT 2013 December 4--5.
All observations were carried out with the GG455 order-blocking filter and the Blue Camera CCD, which imprinted strong fringing redward of about 7000 \AA.
The detector was read out at 400 kHz with 1$\times$1 binning.
Quartz lamp flats and arc lamps for wavelength calibration were taken immediately after 
each science observation at the same position on the sky.
At least one spectrophotometric standard was targeted per night.

All observations are reduced using custom scripts.  Images are bias-subtracted
and corrected for bad pixels.
A normalized flat field is created at the same location as the science trace on the CCD
and is used to remove pixel-to-pixel variations in the science frame, including most (but not all)
of the fringing.
Spectra are then optimally extracted following the method described in \citet{Horne:1986bg}.
Wavelength calibration is carried out by fitting Gaussians to 19 strong emission lines from HgAr
for the arc lamp frames using the 400 l mm$^{-1}$ grating, and 11 emission lines from CuHeAr
for the arc lamp frames using the 1200 l mm$^{-1}$ grating in pixel space.
A fourth order polynomial fit is used to map pixels to wavelengths in an automated fashion for each target.
Finally, the extracted spectrum was divided by a spectrophotometric standard observed on the same night to correct for
wavelength-dependent throughput losses.

\subsection{UH 2.2-m/SuperNova Integral Field Spectrograph}

We acquired low-resolution ($R$$\approx$1300) optical spectra for 40 stars 
on UT 2014 January 19 and 21 with the SuperNova Integral Field Spectrograph (SNIFS) 
at the University of Hawai'i's 88$''$ (2.2 m) telescope located on Maunakea, Hawai'i.
SNIFS is an integral-field spectrograph that uses a microlens array to disperse a 6$''$$\times$6$''$ field of view
into two channels spanning 3200--11000~\AA \ (\citealt{Lantz:2004dk}).
Multiple O/B standards were observed on each night.
After basic image reduction and rectification into data cubes,
each spectrum were extracted and wavelength calibrated with the SNIFS reduction pipeline
(\citealt{Aldering:2006fy}; \citealt{Scalzo:2010ec}).
Details for each target are listed in Table~\ref{tab:specobs}.

\subsection{P60/Robo-AO}{\label{sec:roboao}}

We obtained 1523 adaptive optics images of 1011 targets from our parent sample of 2060 active stars
with Robo-AO at the Palomar 60$''$ (1.5-m) telescope between July 2013 and June 2015.
Robo-AO is an efficient autonomous adaptive optics system that provides diffraction-limited AO observations at optical wavelengths using an 
ultraviolet laser for wavefront sensing (\citealt{Baranec:2013ey}; \citealt{Baranec:2014jc})
and an intelligent queue system for target selection (\citealt{Riddle:2014hh}).

For each observation, Robo-AO's EMCCD camera produces a data cube typically composed of 256 fast readouts
with short exposures.  These frames are combined using a shift-and-add pipeline 
for each observation to produce a final science image with a field of view of 44$''$$\times$44$''$ 
that has been resampled to 21.6 mas pix$^{-1}$, 
or half the native plate scale (see \citealt{Law:2014is} for details).
The plate scale and north orientation are derived from observations of globular clusters
taken on observing runs throughout the same time period as these data.
Because targets tend to be faint and red, most of our observations are carried out with the SDSS $i'$ filter 
with integration times of 30--120 s.
When possible, we obtained multiple observations of candidate visual binaries to test for
common proper motion.
Details about our individual observations can be found in Table~\ref{tab:roboao}.


\begin{figure}
  \vskip -0. in
  \hskip -0.1 in
  \resizebox{3.6in}{!}{\includegraphics{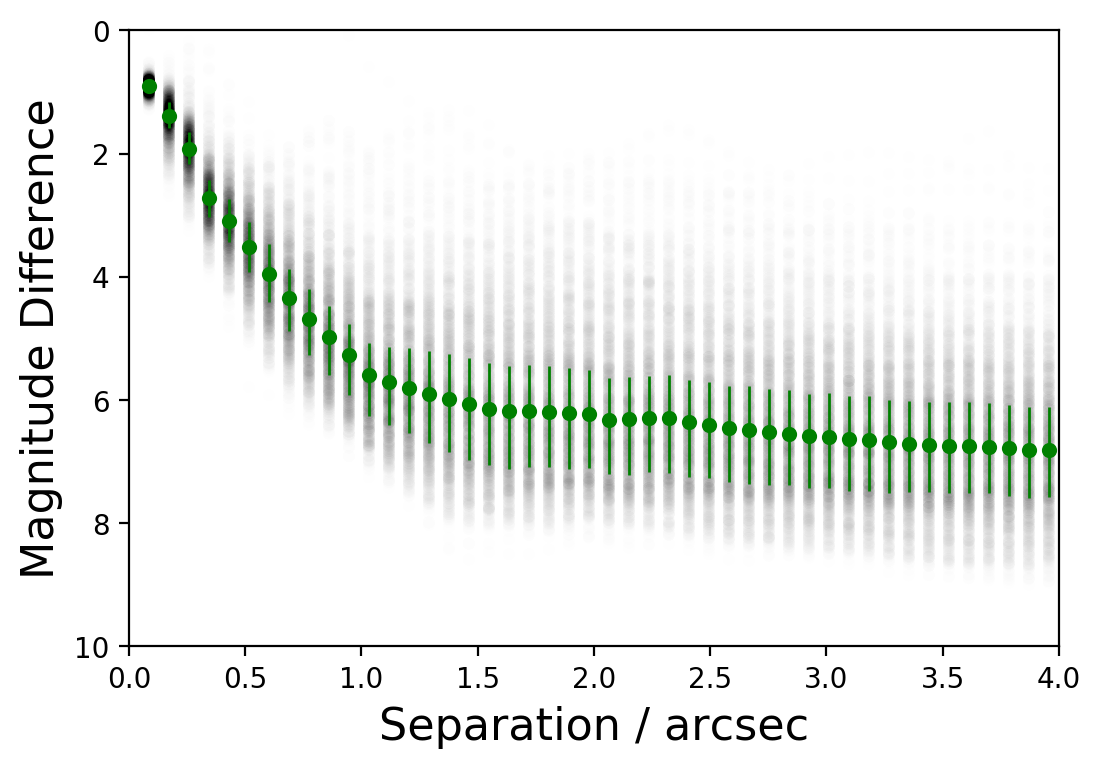}}
  \vskip -.1 in
  \caption{Overview of Robo-AO contrast curves from our observations.  5$\sigma$ sensitivity limits (overlapping gray circles) are derived using
  injection-recovery of each star's PSF.  The median contrasts and upper and lower quartiles are shown in green.   \label{fig:aocontrast} } 
\end{figure}

FWHM values are calculated using the averaged radial profile of the PSF.  When seeing conditions degrade,
the shift-and-add procedure locks on to noise spikes and produces a narrow core in the final image.
For these images, which would otherwise imply sub-diffraction-limited resolution, we ignore the 
inner 5 pix for our FWHM measurement.
The typical FWHM is about 0$\farcs$18, which compares with the diffraction limit of $\approx$0$\farcs$12 at 750 nm.  The median seeing at Palomar Observatory is about 1$\farcs$1.
73\% of our observations have FWHM $<$ 0$\farcs$25 and 
11\% of our observations have FWHM $<$ 0$\farcs$15.  
These measurements are reported in Table~\ref{tab:roboao}.

Image performance metrics and contrast curves are generated for each target following
\citet{Law:2014is} and \citet{Ziegler:2017kq}.  
To summarize, AO correction is assessed using PSF core size.  Targets are divided into high-, medium-, or low-performance groups,
which vary primarily with target brightness and natural seeing conditions.\footnote{Representative contrast curves for
each performance group are as follows: 
High: \{0.7, 1.6, 3.9 , 5.5, 6.3, 6.4, 6.4\} mag,
Medium: \{0.7, 1.6, 3.2 , 4.4, 5.0, 5.0, 5.0\} mag, and
Low: \{0.5, 1.1, 2.2 , 3.1, 3.5, 3.5, 3.5\} mag
at \{0$\farcs$1, 0$\farcs$2, 0$\farcs$5 , 1$''$, 2$''$, 3$''$, 4$''$\}, respectively.}
5$\sigma$ contrast curves are derived using a Monte Carlo injection-recovery analysis of artificial companions generated from the 
primary's PSF.  Contrast curves from our observations are summarized in Figure~\ref{fig:aocontrast}; we typically reach
$\Delta$$i'$ $\approx$ 5 mag at 1$''$.
In Section \ref{sec:vbs} we discuss the visual binaries and fainter candidate companions in our images.

\startlongtable
\begin{deluxetable}{lcccc}
\renewcommand\arraystretch{0.9}
\tabletypesize{\small}
\setlength{ \tabcolsep } {.1cm} 
\tablewidth{0pt}
\tablecolumns{5}
\tablecaption{Robo-AO Observations \label{tab:roboao}}
\tablehead{
       \colhead{2MASS} & \colhead{UT Date}  &  \colhead{}  & \colhead{Exp.} & \colhead{FWHM} \\
       \colhead{ID}    & \colhead{(Y-M-D)}       &  \colhead{Filter}  & \colhead{(s)} & \colhead{($''$)}
        }   
\startdata
         J00055520+4129289    &         2014-08-24    &          SDSS $i'$    &          60    &        0.28    \\
         J00074264+6022543    &         2013-10-25    &            LP600      &         120    &        0.26    \\
         J00074264+6022543    &         2014-11-08    &          SDSS $i'$    &          60    &        0.16    \\
         J00080642+4757025    &         2013-10-25    &          SDSS $i'$    &         120    &        0.20    \\
         J00085391+2050252    &         2013-10-24    &          SDSS $i'$    &         120    &        0.25    \\
         J00085391+2050252    &         2014-11-06    &          SDSS $i'$    &          60    &        0.13    \\
         J00114643--1139553   &         2014-08-28    &          SDSS $i'$    &          60    &        0.26    \\
         J00120761--1550327   &         2014-08-29    &          SDSS $i'$    &          60    &        0.35    \\
         J00133841+5245050    &         2014-08-26    &          SDSS $i'$    &          60    &        0.21    \\
         J00133841+5245050    &         2014-11-06    &          SDSS $i'$    &          60    &        0.13    \\
\multicolumn{5}{c}{$\cdots$} \\
\enddata
\tablecomments{Table 2 is published in its entirety in the machine-readable format.
      A portion is shown here for guidance regarding its form and content.}
\end{deluxetable}

\section{Results}{\label{sec:results}}

\subsection{Spectral Classification}

Spectral types are determined using the \texttt{Hammer} classification package (\citealt{Covey:2007bj}), which
measures a suite of indices and assigns a spectral type by comparing these values
to spectral standards.
\citet{West:2011dx} show that these classifications are generally accurate to $\pm$1 subclass,
but for late-M dwarfs there is an average systematic offset of $\approx$0.4 subtypes 
towards earlier types.
We therefore also assign spectral types using the visual classifying feature in \texttt{Hammer}.
These two methods are generally in agreement, but our visual types are found to be more reliable, so
we adopt an uncertainty of $\pm$0.5 subtypes for these classifications.
As expected from our color cuts, the vast majority of objects for which we obtained
spectra fall between K5 and M5.
Both the automated (index-based) and visual results are reported 
in Table \ref{tab:specobs} together with TiO5 indices, which track the onset and
strengthening of TiO absorption in the emergent spectra of M dwarfs (\citealt{Reid:1995kw}).


\begin{figure}
  \vskip -0.7 in
  \hskip -1 in
  \resizebox{5.6in}{!}{\includegraphics{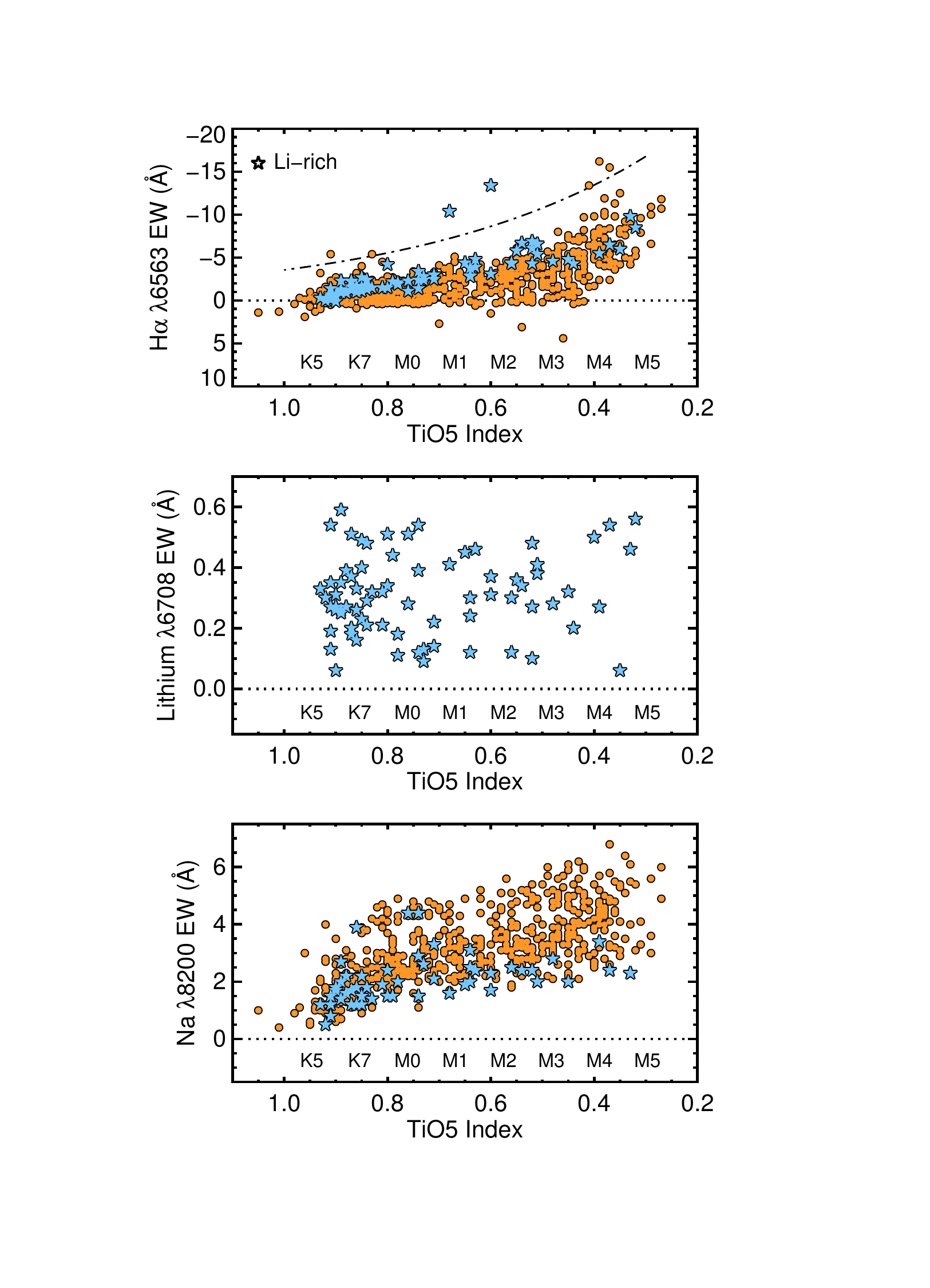}}
  \vskip -0.9 in
  \caption{Age and gravity-dependent line strengths from our low-resolution optical spectra.  
 \emph{Top panel}: H$\alpha$ equivalent width as a function of TiO5 index strength.
 The maximum H$\alpha$ emission from chromospheric activity traces an envelope that
 increases toward larger equivalent widths at later types.  
 The dot-dashed curve represents an approximate boundary between saturated 
 chromospheric emission and emission originating from disk accretion identified by \citet{BarradoyNavascus:2003vz}.
 TiO5 values are converted to spectral types using the relation from \citet{Reid:1995kw}.
 One star, 2MASS J15354856-2958551, has an exceptionally high line strength and lies off the plot.
  Blue stars denote objects with \ion{Li}{1} absorption in their spectra.
  \emph{Middle panel}: \ion{Li}{1} line strength as a function of TiO5 index strength.  A wide range of lithium equivalent widths
  are apparent, implying ages $<$100 Myr for these M dwarfs in our sample.
  \emph{Bottom panel}: Total equivalent width of the gravity-sensitive \ion{Na}{1} doublet at $\approx$8200~\AA.
  Young stars with low surface gravities are expected to have lower sodium strengths.
\label{fig:specew} } 
\end{figure}


\begin{figure*}
  \vskip -0.5 in
  \hskip 0. in
  \resizebox{7.in}{!}{\includegraphics[angle=90]{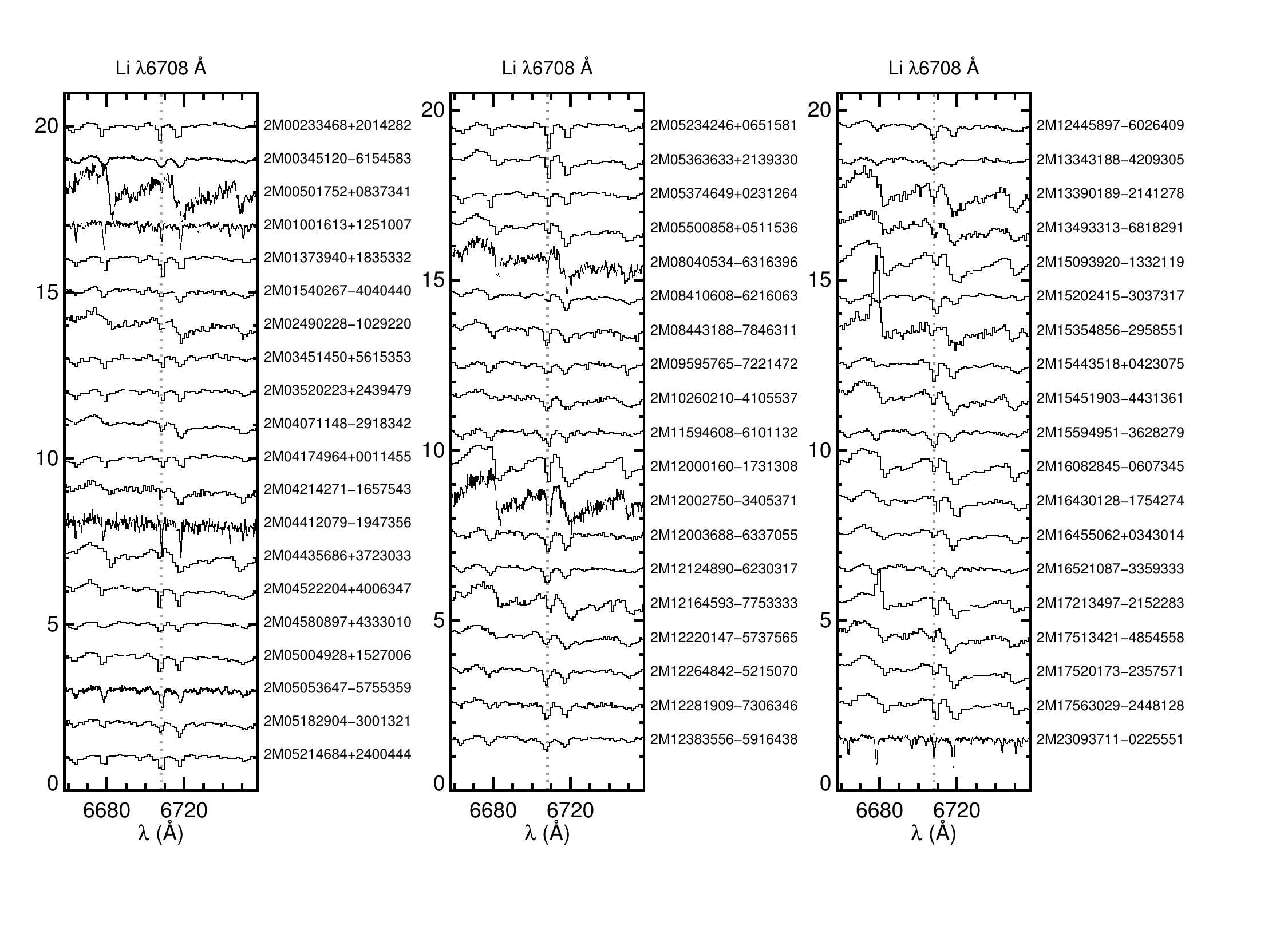}}
  \vskip -.2 in
  \caption{Spectra of 58 lithium-rich stars from the subset of activity-selected targets for which we obtained follow-up spectroscopy.
  The observations were taken with several spectrographs and modes spanning a range of resoling powers from
  $R$$\approx$1300--5900 (details can be found in Table \ref{tab:specobs}).  The \ion{Li}{1} $\lambda$6708 \AA \ line
  is marked with a gray dotted line.
   \label{fig:lithium} } 
\end{figure*}

\subsubsection{H$\alpha$ Emission}

H$\alpha$ emission is observed in the vast majority of our spectra.  We measure equivalent
widths by fitting a Gaussian function centered at 6563 \AA \ using the 
curve-fitting package \texttt{MPFIT} (\citealt{Markwardt:2009wq}) and integrating under the best-fit model.
Each fit was visually inspected to ensure the emission line peak was correctly identified and modeled.
Our threshold for clear line emission is $<$--0.5~\AA.
For equivalent widths between 0.0 to --0.5~\AA, the emission is either very weak or questionable
based on visual inspection.  Values in this range are less reliable because of the low resolving power 
of our data and should be treated with caution.  High-resolution spectra may be needed to unambiguously 
search for H$\alpha$ emission in those stars.
Equivalent widths $>$0~\AA \ indicate that H$\alpha$ is seen in absorption.
H$\alpha$ line strengths are listed in Table \ref{tab:specobs}.
Uncertainties are determined by comparing equivalent widths of the same targets
on the same night; we estimate errors of 20\% for the quoted line strengths.


\begin{figure}[t!]
  \vskip -0.1 in
  \hskip -0.3 in
  \resizebox{4in}{!}{\includegraphics{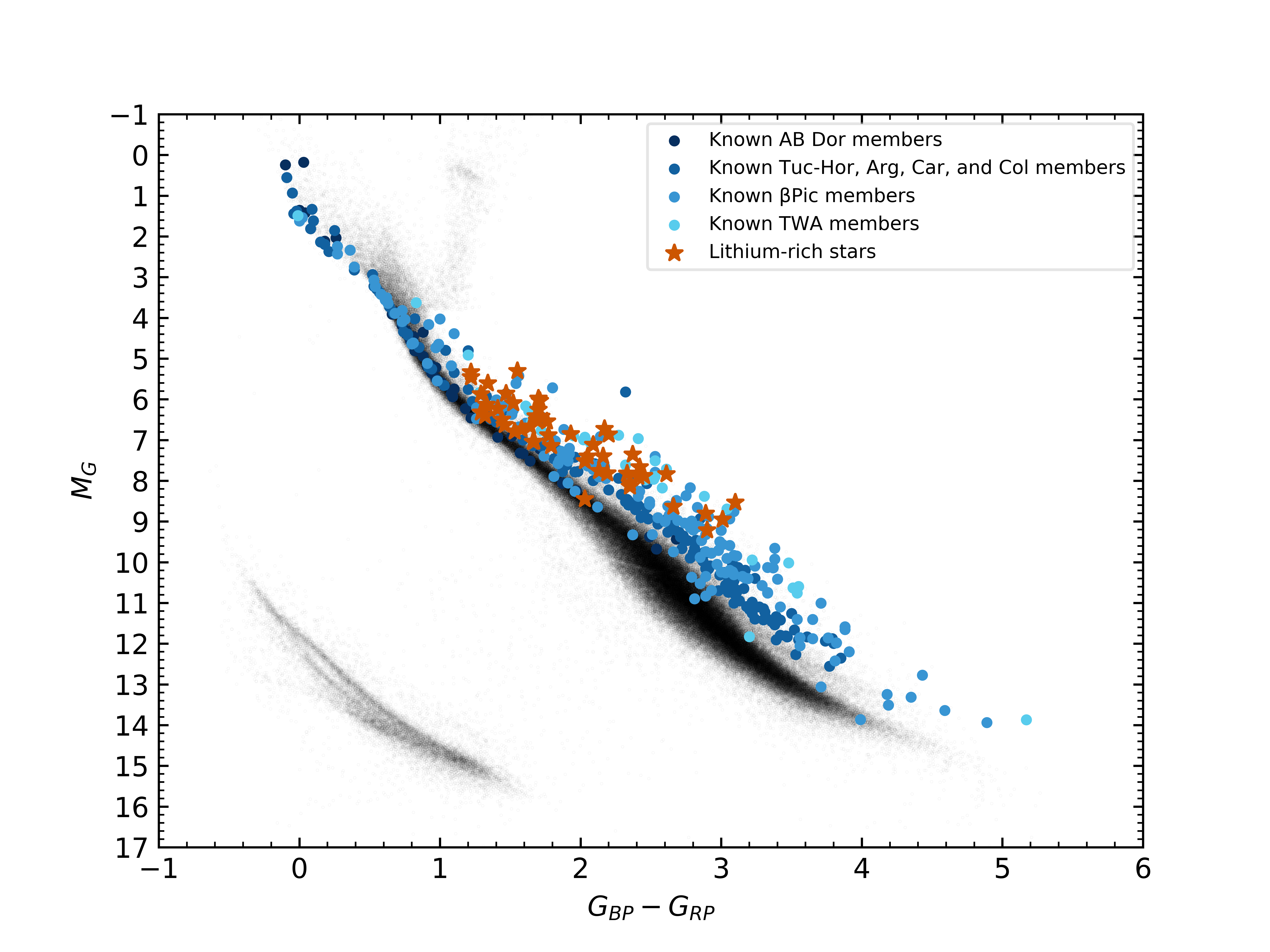}}
  \vskip -.2 in 
  \caption{Positions of lithium-rich stars (red stars) in the $Gaia$ color-magnitude diagram relative to known moving group
  members from \citet{Malo:2013gn}.  The $Gaia$ CMD shows stars within 100 pc with spurious entries removed 
  following \citet{Lindegren:2018gy}.
   \label{fig:gaiacmd} } 
  \vskip -.1 in 
\end{figure}

H$\alpha$ equivalent widths are shown as a function of TiO5 index strength in Figure~\ref{fig:specew}.
Our targets trace an envelope of H$\alpha$ emission that increases in strength from 
about --3~\AA \ at TiO5 values of 0.9 ($\approx$K7) to $>$10~\AA \ at TiO5 values of 0.4 ($\approx$M4).
The shape of this envelope bears a close resemblance to other large spectroscopic 
samples of  M dwarfs (e.g., \citealt{Riaz:2006du}; \citealt{Gaidos:2014if}; \citealt{Kraus:2014ur}).
\citet{BarradoyNavascus:2003vz} identify an empirical division that separates 
accretion-induced H$\alpha$ emission from saturated chromospheric activity.
Eight stars have exceptionally strong H$\alpha$ emission that
falls on or above the saturated chromospheric curve in Figure~\ref{fig:specew}
and may originate in part from disk accretion:
2MASS J10260210--4105537,  2MASS J13314666+2916368, 
2MASS J13573397--3139105, 2MASS J14255593+1412101, 2MASS J15354856--2958551, 
2MASS J17213497--2152283, 2MASS J18464675+0043260, and 2MASS J19300396--2939322.

\subsubsection{Lithium}

\ion{Li}{1} $\lambda$6708~\AA \ absorption is a well-established indicator of youth
 in the atmospheres of low-mass stars (e.g., \citealt{Soderblom:2014ve}).  
 Lithium burning occurs in stellar cores through proton capture reactions
 at temperatures of about 2.5$\times$10$^6$ K,
and the depletion of lithium among late-type stars with partially or fully convective envelopes 
is a strong function of both mass and age 
(\citealt{Basri:1996aa}; \citealt{Chabrier:1996kc}; \citealt{Bildsten:1997cx}).  
The presence and strength of lithium therefore acts as a sensitive chronometer 
for masses between about 0.06--0.6~\Msun.

Lithium is apparent in 58 stars from the subset of our parent sample for which we obtained spectra (632 out of 2060 stars;
see Tables \ref{tab:specobs} and \ref{tab:lithiumstars}).
Line profiles are fit with Gaussian functions to calculate equivalent widths.
We estimate uncertainties of about 20\% based on multiple measurements of the same targets in our sample.
Our low-resolution observations are shown in Figure~\ref{fig:lithium} and are sensitive to the strongest lines, 
so there are likely to be additional stars with weaker levels of lithium below our detection limits (about 50--200 m\AA)
that we were not sensitive to.

Equivalent widths range from $\approx$100--600 m\AA \ and span the full range of spectral types from K5 to M5 (middle panel; Figure~\ref{fig:specew}).
The diversity of line strengths implies a range of ages for these stars,
with the highest equivalent widths corresponding to ages at least as young as TWA ($\approx$10 Myr)
based on empirical lithium depletion boundaries for young clusters
 (e.g., \citealt{Neuhauser:1997jg}; \citealt{Mentuch:2008gb}).
For spectral types $>$K7, all of our stars exhibiting lithium are expected to have ages younger than
the Pleaides ($\approx$125~Myr; \citealt{Stauffer:1998kt}).
We note that our lithium stars tend to have high H$\alpha$ emission line strengths (top panel; Figure~\ref{fig:specew})
and the lower sodium values (lower panel), pointing to higher magnetic activity levels, larger physical radii, 
and lower surface gravities.


\begin{figure*}
  \vskip 0.1 in
  \hskip .3 in
  \resizebox{6.5in}{!}{\includegraphics{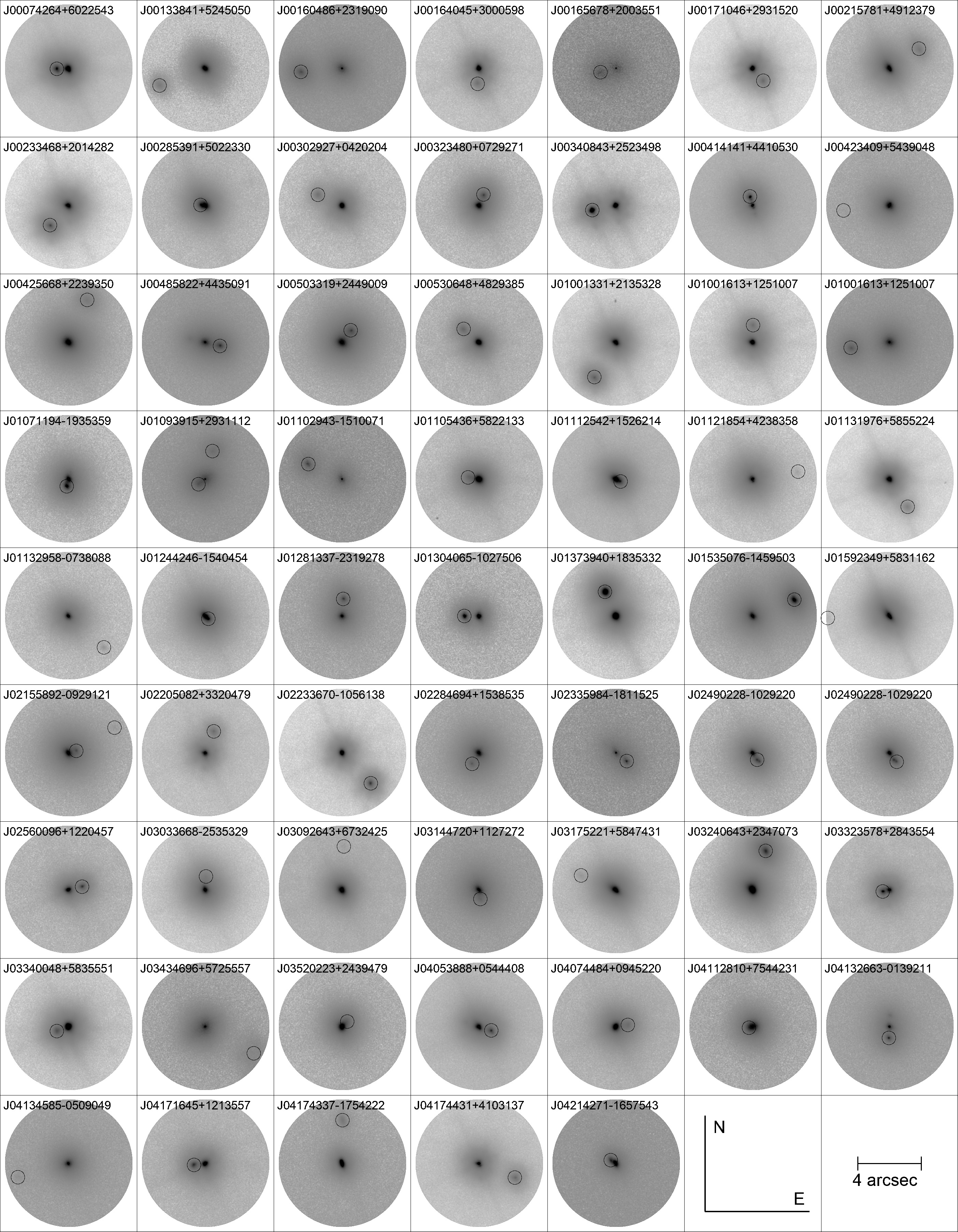}}
  \vskip 0 in
  \caption{Point sources identified in our Robo-AO observations for R.A.s between 00 h--04 h.  Circles mark the locations of point sources.
  The field of view of each image is 8$''$$\times$8$''$.  The sky orientation is denoted at the bottom right of the figure.   
   \label{fig:roboao1} } 
\end{figure*}


\begin{figure*}
  \vskip 0.1 in
  \hskip 0.3 in
  \resizebox{6.5in}{!}{\includegraphics{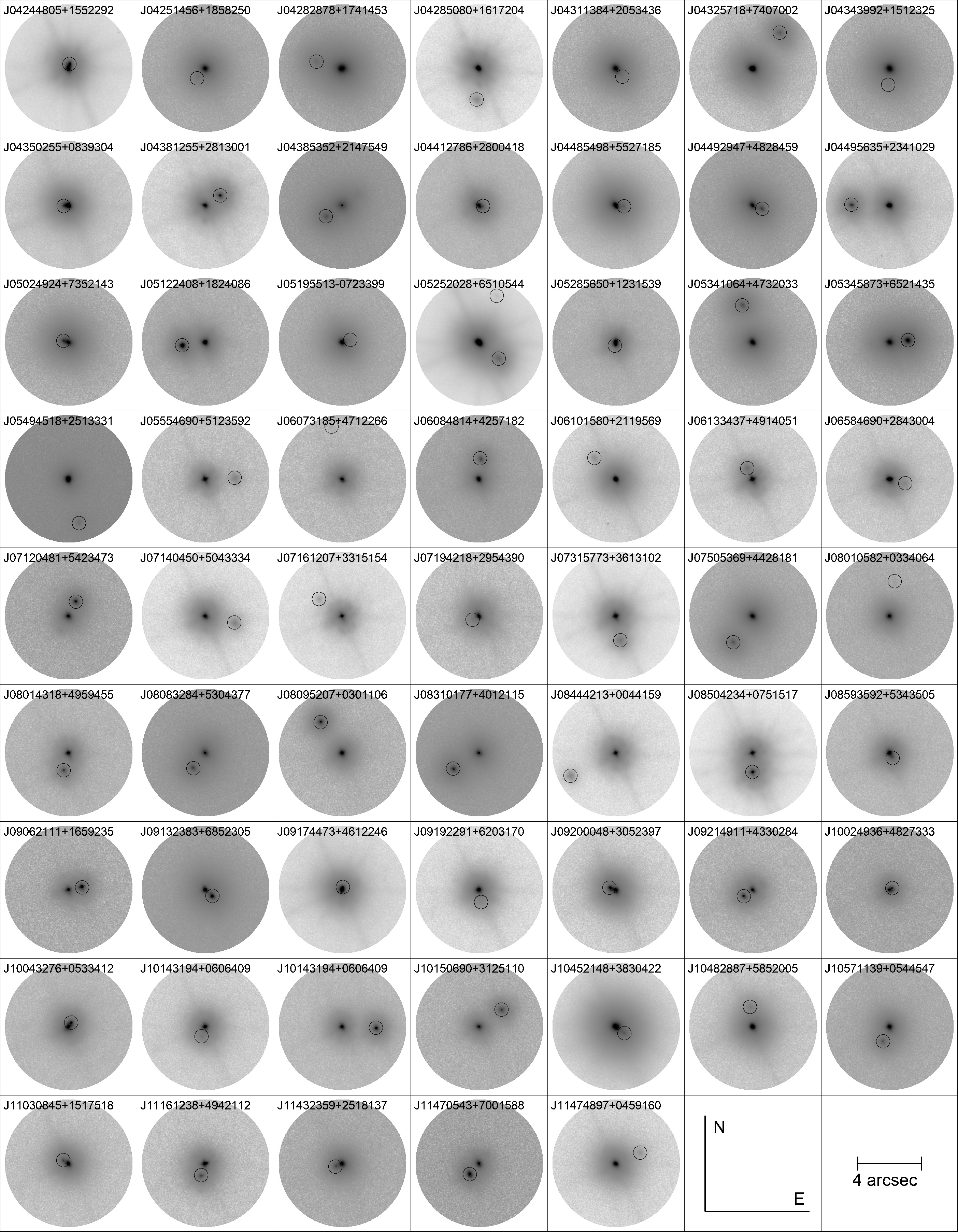}}
  \vskip 0 in
  \caption{Same as Figure~\ref{fig:roboao1} but for R.A.s between 04--11 h.
   \label{fig:roboao2} } 
\end{figure*}


\begin{figure*}
  \vskip 0.1 in
  \hskip 0.3 in
  \resizebox{6.5in}{!}{\includegraphics{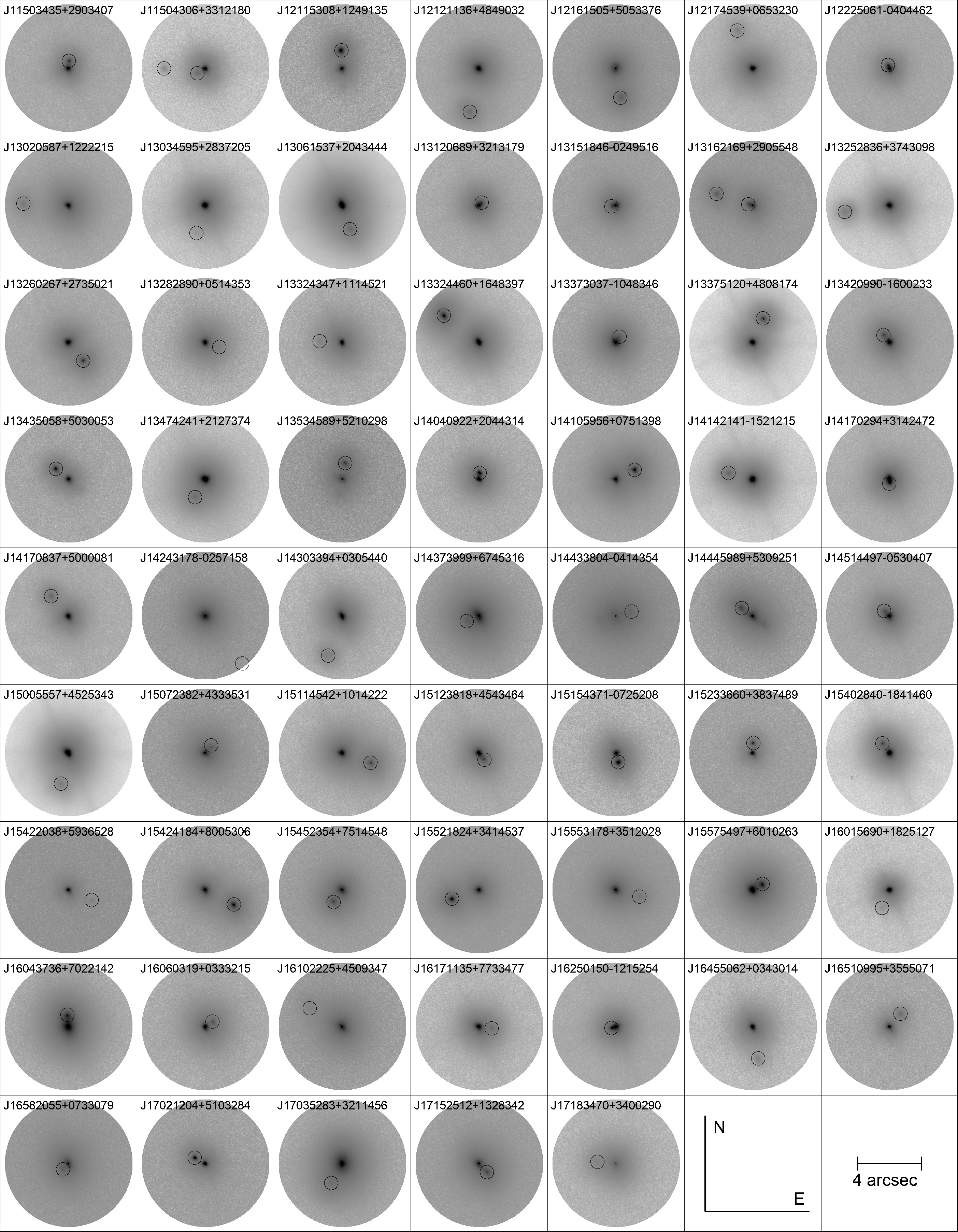}}
  \vskip 0 in
  \caption{Same as Figure~\ref{fig:roboao1} but for R.A.s between 11--17 h.
   \label{fig:roboao3} } 
\end{figure*}


\begin{figure*}
  \vskip 0.1 in
  \hskip 0.3 in
  \resizebox{6.5in}{!}{\includegraphics{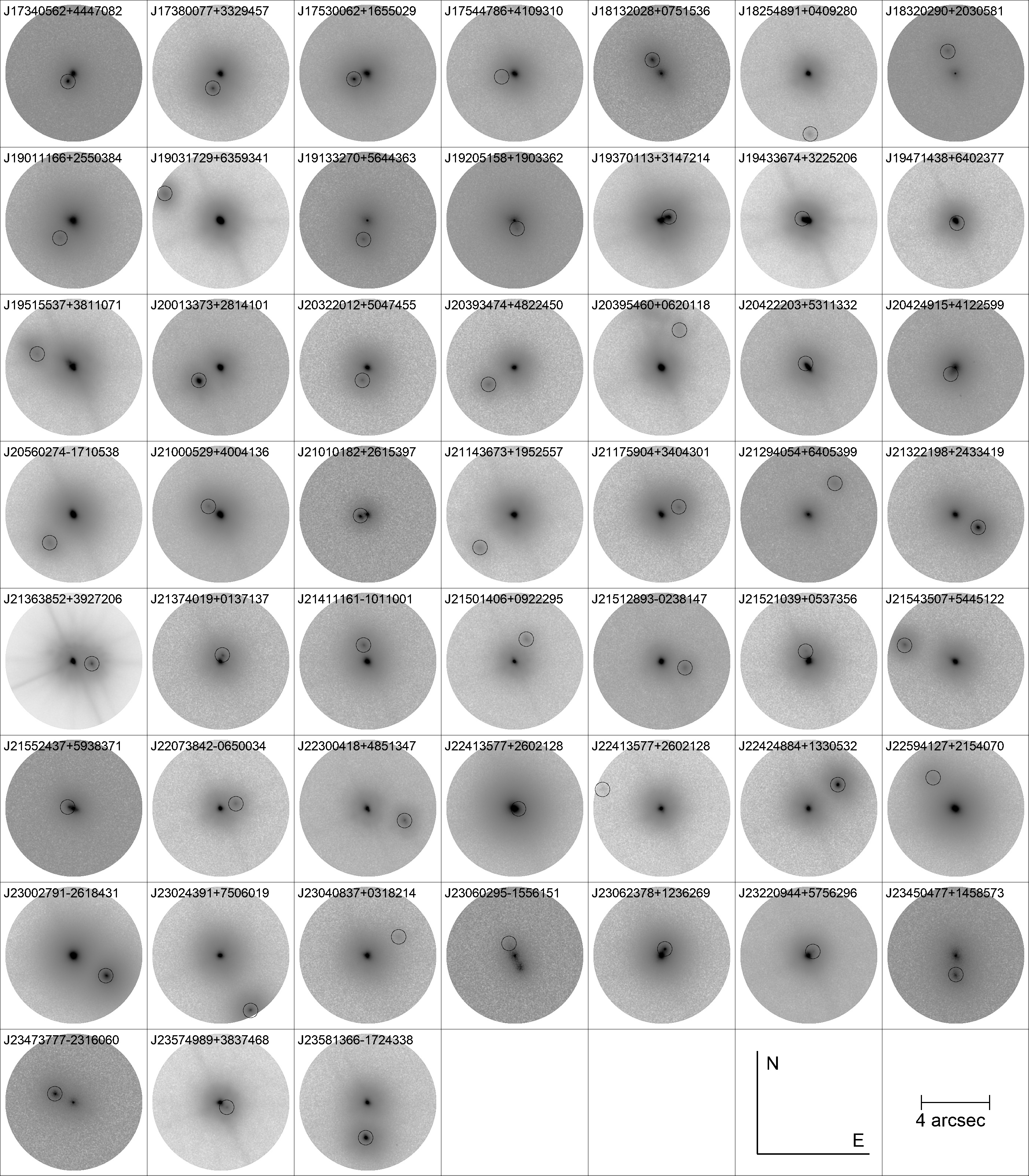}}
  \vskip 0 in
  \caption{Same as Figure~\ref{fig:roboao1} but for R.A.s between 17--00 h.
   \label{fig:roboao4} } 
\end{figure*}

 \begin{longrotatetable}
\startlongtable
\begin{deluxetable*}{lccccccccccc}
\renewcommand\arraystretch{0.9}
\tabletypesize{\footnotesize}
\setlength{ \tabcolsep }{.1cm}
\tablewidth{0pt}
\tablecolumns{12}
\tablecaption{Properties of Lithium-Rich Stars\label{tab:lithiumstars}}
\tablehead{
       \colhead{2MASS} &        & \colhead{SpT}  & \colhead{RV}  & \colhead{RV}  & \colhead{$\mu_{\alpha} \cos(\delta)$\tablenotemark{a}}  & \colhead{$\mu_{\delta}$\tablenotemark{a}}   & \colhead{Distance\tablenotemark{a}} & \colhead{BANYAN $\Sigma$} & \colhead{Literature} & \colhead{Adopted} & \colhead{YMG} \\
       \colhead{Name} & \colhead{SpT}   &  \colhead{Ref.}  & \colhead{(km/s)}   & \colhead{Ref.}  & \colhead{(mas yr$^{-1}$)}    & \colhead{(mas yr$^{-1}$)}  & \colhead{(pc)}      & \colhead{Best Hyp.} & \colhead{YMG} & \colhead{YMG} & \colhead{Ref.} 
  }
  \startdata
J00233468+2014282     & M0   &  TW  & --2.2 $\pm$ 0.6  & Sh17     &      65.97   $\pm$      0.10                  &    --37.38    $\pm$      0.11                    &      62.89  $\pm$       0.25                  &  Field            & $\beta$ Pic        &  $\beta$ Pic   &          Le09, Ma13, Ma14a, Sh17 \\ 
J00345120--6154583    & K7   &  TW  &    11 $\pm$ 5    & Kr14     &      88.69   $\pm$      0.04                  &    --52.66    $\pm$      0.04                    &      44.50  $\pm$       0.05                  &  Tuc-Hor          & Tuc-Hor            &  Tuc-Hor       &          Zu00, Zu01a, To08, Kr14      \\ 
J00501752+0837341     & M5   &  TW  &  2.15 $\pm$ 2.0  & Sh17     &      68.02   $\pm$      0.26                  &    --35.05    $\pm$      0.13                    &      60.86  $\pm$       0.52                  &  Field            & $\beta$ Pic        &  $\beta$ Pic   &          Sh17     \\                      
J01001613+1251007     & K5   &  TW  &     $\cdots$     & $\cdots$ &      47.41   $\pm$      0.08                  &    --31.57    $\pm$      0.06                    &      94.15  $\pm$       0.41                  &  Field            & $\cdots$           &  Field         &          $\cdots$    \\
\enddata
\end{deluxetable*}
\end{longrotatetable}

Figure~\ref{fig:gaiacmd} shows the position of our lithium-rich stars in the $Gaia$ color-magnitude diagram (CMD) relative 
to known members of young moving groups.  
The $Gaia$ DR2 CMD is constructed largely adhering to recommendations by \citet{Lindegren:2018gy} with the following additional
restrictions: parallaxes $>$10 mas; parallax SNR $>$10; and photometric SNR $>$10 in $G$, $G_B$, and $G_R$ 
bandpasses.
Most of the lithium stars lie above the main sequence and are consistent with
the isochrones traced out by AB Dor ($\approx$120 Myr); Tuc-Hor, Argus, Carina, and Columba ($\approx$40--50 Myr);
$\beta$ Pic ($\approx$23 Myr); and TWA ($\approx$10 Myr).

\subsubsection{Sodium}

Like other alkali elements, the relative strength of the \ion{Na}{1} doublet at 8183~\AA \ and 8195~\AA \ is sensitive to 
atmospheric pressure and surface gravity (e.g., \citealt{Slesnick:2006ij}).   
\citet{Schlieder:2012iw} show that this doublet can act as a useful tracer of youth for spectral types $>$M3
because of its prominence relative to the pseudo-continuum at cool temperatures and its stronger dependence on surface gravity 
 at lower masses.
We simultaneously fit two Gaussians to these neighboring lines and report the total equivalent width of the pair
for each spectrum in Table \ref{tab:specobs}.
The bottom panel of Figure~\ref{fig:specew} shows a general strengthening of the lines at lower temperatures with significant
spread for a given spectral type.  
Beyond M0, stars with lithium tend to lie near the lower envelope of our sodium measurements, in agreement with
the expectation of large radii, low surface gravities, and young ages for these objects.

 \startlongtable
\begin{deluxetable*}{lcccccccc}
\renewcommand\arraystretch{0.9}
\tabletypesize{\small}
\setlength{ \tabcolsep }{.1cm} 
\tablewidth{0pt}
\tablecolumns{9}
\tablecaption{Robo-AO Point Source Detections \label{tab:vbs}}
\tablehead{
     \colhead{2MASS} & \colhead{UT Date}       &  \colhead{}        & \colhead{Candidate} & \colhead{} & \colhead{Sep}    & \colhead{P.A.}       & \colhead{Significance} & \colhead{Performance} \\
     \colhead{ID}    & \colhead{(Y-M-D)}       &  \colhead{Filter}  & \colhead{Name}      & \colhead{$\Delta$mag}        & \colhead{($''$)} & \colhead{($\degr$)}  & \colhead{($\sigma$)}   & \colhead{Metric\tablenotemark{a}} 
        }   
\startdata
 J00074264+6022543    &         2014-11-08    &            $i'$  &    CC1     &        0.30 $\pm$  0.05     &        0.86 $\pm$  0.03     &        94 $\pm$  2   &        12.5     &         H    \\
 J00133841+5245050    &         2014-08-26    &            $i'$  &    CC1     &        2.76 $\pm$  0.03     &        3.13 $\pm$  0.03     &       110 $\pm$  3   &        11.5     &         H    \\
 J00133841+5245050    &         2014-11-06    &            $i'$  &    CC1     &        2.39 $\pm$  0.05     &        3.16 $\pm$  0.04     &       109 $\pm$  2   &         7.2     &         M    \\
 J00133841+5245050    &         2014-11-06    &            $i'$  &    CC1     &        2.74 $\pm$  0.02     &        3.16 $\pm$  0.04     &       109 $\pm$  3   &         6.4     &         H    \\
 J00133841+5245050    &         2014-11-06    &            $i'$  &    CC2     &        3.41 $\pm$  0.04     &        0.86 $\pm$  0.03     &        94 $\pm$  2   &         5.9     &         M    \\
 J00160486+2319090    &         2014-11-06    &            $i'$  &    CC1     &        0.41 $\pm$  0.05     &        2.66 $\pm$  0.05     &        94 $\pm$  2   &         9.6     &         M    \\
\enddata
\tablenotetext{a}{Image performance metrics assessed from the PSF core size.  These are divided into three groups: high-performance (``H''), medium-performaance (``M''), and low-performance (``L'').
Typical contrasts for each group are described in detail in Section~\ref{sec:roboao}.}
\end{deluxetable*}

\subsection{Visual Binaries}{\label{sec:vbs}}

Point sources are identified in our Robo-AO images following the procedure described in \citet{Ziegler:2017kq}.
Each image is inspected visually and through an automated source-finding algorithm.
We focus on  a 4$''$ radius surrounding each star where our AO observations are most advantageous compared to all-sky 
seeing-limited surveys.
Altogether point sources are found near 239 stars with optical contrasts peaking at $\Delta$mag$\approx$0.5 and
reaching $\Delta$mag$\approx$7 for the faintest objects identified (Figures~\ref{fig:roboao1}--\ref{fig:roboao4}).
Relative contrasts are measured using aperture photometry at wide separations and processed images of the companion
after PSF subtraction at close separations.  
Uncertainties in astrometry and contrast are estimated based on systematic errors caused by blending and due to maximum 
orientation changes during the observing period, which we estimate to be $\pm$1$\fdg$5 using calibration fields.
Contrasts, separations, position angles, and estimated uncertainties for all point sources can be found in Table~\ref{tab:vbs}.
Note that targets reported in our complete list of observations in Table~\ref{tab:roboao} that do not have nearby point sources listed in Table~\ref{tab:vbs}
imply that they are single, at least to within our sensitivity limits.

In Table~\ref{tab:cpm} we compare the observed and expected astrometry for candidate binary companions with
multi-epoch imaging to test whether they are background stars or physically bound systems.
$\chi^2$ values for the common proper motion ($\chi^2_\mathrm{CPM}$) 
and background ($\chi^2_\mathrm{BG}$) hypotheses are calculated as follows:

\begin{displaymath}
\chi^2_\mathrm{CPM} = \sum_{i=1}^{N-1} \left( \frac{(\theta_{\mathrm{meas}, i} - \theta_{\mathrm{ref}, i})^2}{\sigma_{\theta, \mathrm{meas}, i}^2 + \sigma_{\theta, \mathrm{ref}, i}^2} + 
 \frac{(\rho_{\mathrm{meas}, i} - \rho_{\mathrm{pred}, i})^2}{\sigma_{\rho, \mathrm{meas}, i}^2 + \sigma_{\rho, \mathrm{ref}, i}^2} \right),
\end{displaymath}

\begin{displaymath}
\chi^2_\mathrm{BG} = \sum_{i=1}^{N-1} \left( \frac{(\theta_{\mathrm{meas}, i} - \theta_{\mathrm{pred}, i})^2}{\sigma_{\theta, \mathrm{meas}, i}^2 + \sigma_{\theta, \mathrm{pred}, i}^2} + 
 \frac{(\rho_{\mathrm{meas}, i} - \rho_{\mathrm{pred}, i})^2}{\sigma_{\rho, \mathrm{meas}, i}^2 + \sigma_{\rho, \mathrm{pred}, i}^2} \right).
\end{displaymath}

\noindent Here $\theta_{\mathrm{meas}, i}$, $\rho_{\mathrm{meas}, i}$,  $\sigma_{\theta, \mathrm{meas}, i}$, and $\sigma_{\rho, \mathrm{meas}, i}$ are the  measured P.A.,
separation, and their respective uncertainties for epoch $i$ of $N$ total epochs; $\theta_{\mathrm{ref}, i}$, $\rho_{\mathrm{ref}, i}$,  $\sigma_{\theta, \mathrm{ref}, i}$, and $\sigma_{\rho, \mathrm{ref}, i}$ are the same for the reference epoch (here taken to be our first observation of the system); and $\theta_{\mathrm{pred}, i}$, $\rho_{\mathrm{pred}, i}$,  $\sigma_{\theta, \mathrm{pred}, i}$, and $\sigma_{\rho, \mathrm{pred}, i}$ are the predicted relative astrometry of a stationary background source based on the distance, proper motion,
and sky position of the target (Table \ref{tab:vbymg}). 

The Bayesian Information Criterion (BIC; \citealt{Schwarz:1978uv}) is used to assess the 
significance of evidence for or against the background and comoving
models.  It is constructed to reward better fits but penalize more complex models as follows:

\begin{displaymath}
\mathrm{BIC} = \chi^2 + k \ln n
\end{displaymath}

\noindent where $k$ is the number of free parameters in the model and $n$ is the number of epochs.
Lower BIC values are preferred.
Differenced BIC values ($\Delta$BIC = BIC$_\mathrm{BG}$ -- BIC$_\mathrm{CPM}$) for both bound and unbound scenarios are listed in
Table~\ref{tab:cpm}.  Following \citet{Kass:1995aa}, $\Delta$BIC values between 0--3 are interpreted as modest evidence in favor of
common proper motion, $\Delta$BIC values greater than 3 suggest strong evidence for common proper motion,
$\Delta$BIC values between 0 and --3 point to modest evidence in favor of the background model, and 
$\Delta$BIC values less than --3 imply strong evidence for the background model.

\startlongtable
\begin{deluxetable*}{lcccccccccccc}
\renewcommand\arraystretch{0.9}
\tabletypesize{\footnotesize}
\setlength{ \tabcolsep }{.1cm} 
\tablewidth{0pt}
\tablecolumns{13}
\tablecaption{Binary Common Proper Motion Tests\label{tab:cpm}}
\tablehead{
       \colhead{2MASS} & \colhead{$N_\mathrm{epochs}$}  & \colhead{$\Delta t$}  & \colhead{Cand.} & \colhead{$\chi^2_{\nu}$} &  \colhead{$\chi^2_{\nu}$} & \colhead{$\nu$} & \colhead{BIC}  & \colhead{BIC}   & \colhead{$\Delta$BIC} & \colhead{Comp?\tablenotemark{a}} & \colhead{Known}  &  \colhead{Binary}  \\
       \colhead{Name}  & \colhead{}                     & \colhead{(yr)}        & \colhead{Comp.} & \colhead{(BG)}           & \colhead{(CPM)}           & \colhead{}      & \colhead{(BG)} & \colhead{(CPM)} & \colhead{}            & \colhead{}      & \colhead{Binary?}  &  \colhead{Ref.} 
  }
\startdata
J00074264+6022543      &    1 & $\cdots$   & CC1 &   $\cdots$ &  $\cdots$   &    0 & $\cdots$  & $\cdots$ &   $\cdots$  &  SE    & Y  &  Ja14b, Bo15b   \\
J00133841+5245050      &    3 & 0.197      & CC1 &   0.352    &  0.426      &    2 & 6.20      & 6.35     &   -0.148    &  BG?	 & N  &  $\cdots$   \\
J00133841+5245050      &    1 & $\cdots$   & CC2 &   $\cdots$ &  $\cdots$   &    0 & $\cdots$  & $\cdots$ &   $\cdots$  &  SE	 & N  &  $\cdots$   \\
J00160486+2319090      &    1 & $\cdots$   & CC1 &   $\cdots$ &  $\cdots$   &    0 & $\cdots$  & $\cdots$ &   $\cdots$  &  SE	 & N  &  $\cdots$   \\
J00164045+3000598      &    1 & $\cdots$   & CC1 &   $\cdots$ &  $\cdots$   &    0 & $\cdots$  & $\cdots$ &   $\cdots$  &  SE	 & N  &  $\cdots$   \\
J00165678+2003551      &    1 & $\cdots$   & CC1 &   $\cdots$ &  $\cdots$   &    0 & $\cdots$  & $\cdots$ &   $\cdots$  &  SE	 & N  &  $\cdots$   \\
J00171046+2931520      &    2 & 0.847      & CC1 &   0.307    &  0.293      &    1 & 3.77      & 3.76     &   0.0133    &  CPM?	 & N  &  $\cdots$   \\
J00215781+4912379      &    4 & 0.828      & CC1 &   14.1     &  0.795      &    3 & 49.2      & 9.32     &   39.9      &  CPM	 & Y  &  Bo15b   \\
\enddata
\tablenotetext{a}{Status of companion based on differenced Bayesian Information Criterion for the common proper motion and background hypotheses.  
``SE''=Single Epoch; ``BG''=background; ``CPM''=Common Proper Motion.  See Section \ref{sec:vbs} for details.}
\tablerefs{
Beu04=\citet{Beuzit:2004fq},
Ber10=\citet{Bergfors:2010hm}, 
Bo15a=\citet{Bowler:2015ch}, 
Bo15b=\citet{Bowler:2015ja}, 
Dae07=\citet{Daemgen:2007fg}, 
De99=\citet{Delfosse:1999aa}, 
Gu05=\citet{Guenther:2005gu}, 
Ja12=\citet{Janson:2012dc}, 
Ja14b=\citet{Janson:2014fx},
Jay01=\citet{Jayawardhana:2001if}, 
Jo13=\citet{Jodar:2013ja}, 
La08=\citet{Law:2008fn},
Mc01=\citet{McCarthy:2001gt}, 
Sc98=\citet{Schneider:1998aa}, 
WDS=\citet{Mason:2001ff}.
}
\end{deluxetable*}


Altogether 252 sources are detected within 4$''$ of 239 stars.  A single epoch was acquired for most 
candidate companions so some sources may be background stars, but the vast majority are expected
to be physical binaries based on the low number density of comparably bright stars.
We carried out a literature search primarily consulting the Washington Double Star Catalog 
(\citealt{Mason:2001ff}) and identified 88 previously known binaries --- most of which have undergone
significant orbital motion since their discovery -- while the rest appear to be new.

 \subsection{Young Moving Group Members}

Among the 58 lithium-rich stars in our sample, 35 are previously known or suspected members of  young
moving groups or nearby star-forming regions (Table~\ref{tab:lithiumstars}).   Similarly, 51 out of 238 visual binaries 
have been identified as known or candidate YMG members in the literature (Table \ref{tab:vbymg}). 
We used the BANYAN-$\Sigma$ tool from \citet{Gagne:2018wf} to search for additional YMG members
in our our lithium-rich stars and active binary samples.  BANYAN-$\Sigma$ is a Bayesian classifier 
that uses kinematic information to 
determine an object's membership probability for young moving groups within 150 pc.
Compared to previous versions of BANYAN (\citealt{Malo:2013gn}; \citealt{Gagne:2014gp}), this updated package 
uses a refined model of the galactic disk together with spatial and kinematic 
constraints for 27 associations with ages $\lesssim$800 Myr, including nearby star-forming regions
and intermediate-age open clusters.

Results from the BANYAN analysis using default parameters for association locations and space motions
are listed in Tables~\ref{tab:lithiumstars} and \ref{tab:vbymg}.
When available, radial velocities from the literature have been used for the lithium-rich sample.
We do not make use of an instantaneous radial velocity measurement for the active binaries since long-baseline monitoring is needed to
robustly measure the pair's systemic velocity.
The best hypothesis refers to the most probable kinematic and spatial model, including the field.
Results from BANYAN broadly agree with previous assessments, but in many cases either identifies the
field as the most likely hypothesis or disagrees on the most likely moving group.
Altogether an additional seven and ten systems are identified as new candidate moving group members 
from the lithium-rich and active binary samples, respectively.  

\section{Notes on Individual Objects}

Below we comment on new candidate YMG members,
noteworthy individual systems with unusually high
H$\alpha$ emission,  and objects with discrepancies between BANYAN-$\Sigma$ and literature YMG assessments 
from Tables~\ref{tab:lithiumstars} and \ref{tab:vbymg}.
Our final adopted membership status takes into account lithium line strength, $UVW$ kinematics, spatial position, sky position,  
and CMD position when possible.

\textbf{2MASS J00233468+2014282} ---
\citet{Lepine:2009ey} first identify this star as a member of $\beta$ Pic, 
which has been bolstered by several additional studies (\citealt{Malo:2013gn}; \citealt{Malo:2014dk}; \citealt{Shkolnik:2017ex}).
However, the best hypothesis from BANYAN-$\Sigma$ is the field population.
We measure modest lithium absorption ($EW$$\approx$260 m\AA), weak H$\alpha$ emission ($EW$=--1 \AA), and a  
spectral type of M0 from our Mayall spectrum, in close agreement with previous measurements. 
The observed lithium strength is typical of $\beta$ Pic members of this spectral type.
This target is identified as a 1$\farcs$7 binary in the Washington Double Star catalog; we easily recovered this companion
with our Robo-AO observations.  
Using the measured RV of --2.2 $\pm$ 0.6 km s$^{-1}$ from \citet{Shkolnik:2017ex} together with $Gaia$ DR2 astrometry, 
the space velocities of this system are $U$=--11.81 $\pm$ 0.19 km s$^{-1}$, $V$=--17.4 $\pm$ 0.4 km s$^{-1}$, and 
$W$=--8.8 $\pm$ 0.4 km s$^{-1}$.  
This is only 1.3 $\sigma$ (2.3 $\pm$ 1.7 km s$^{-1}$) 
from the locus of $\beta$ Pic members from \citet{Torres:2008vq}\footnote{Differential velocities  
($\Delta$$v$) and uncertainties ($\sigma_{\Delta v}$) are calculated
as follows: $\Delta$$v$ = $\sqrt{(U - U_0)^2 + (V - V_0)^2 + (W - W_0)^2}$, 
$\sigma_{\Delta v}$ = $\sqrt{ (U - U_0)^2  (\sigma_U^2 + \sigma_{U_0}^2) + (V - V_0)^2 (\sigma_V^2 + \sigma_{V_0}^2) + (W - W_0)^2 (\sigma_W^2 + \sigma_{W_0}^2)}$ / $\Delta$$v$}.  
Given the excellent agreement of this system with other established $\beta$ Pic members,
we adopt previous membership assessments in this group over BANYAN's field hypothesis 

\textbf{2MASS J00501752+0837341} ---
This M5 star was proposed as a $\beta$ Pic member by \citet{Shkolnik:2017ex},
who also identify it as an SB2, but the best hypothesis from BANYAN-$\Sigma$ is the field.
We measure a lithium EW of $\approx$60 m\AA \ and strong H$\alpha$ emission.
Using the measured RV of 2.15 $\pm$ 2.0 km s$^{-1}$ from \citet{Shkolnik:2017ex} together with $Gaia$ DR2 astrometry, 
the space velocities of this system are $U$=--12.7 $\pm$ 0.6 km s$^{-1}$, $V$=--16.6 $\pm$ 1.0 km s$^{-1}$, and 
$W$=--7.8 $\pm$ 1.6 km s$^{-1}$.  This is 1.5 $\sigma$ (3.0 $\pm$ 2.0 km s$^{-1}$) 
from the locus of $\beta$ Pic members from \citet{Torres:2008vq}.
Given the good agreement with known members,
we adopt previous membership assessment in $\beta$ Pic over BANYAN's field hypothesis

\textbf{2MASS J01540267--4040440} ---
This K7 star was proposed as a Columba member by \citet{Malo:2014dk}, but the best hypothesis from
BANYAN-$\Sigma$ is the field.
We measure lithium absorption with a depth of $\approx$160~m\AA \ from our SOAR/Goodman data.
Using the measured RV of 12.7 $\pm$ 0.2 km s$^{-1}$ from \citet{Malo:2014dk} together with $Gaia$ DR2 astrometry, 
the space velocities of this system are $U$=--11.42 $\pm$ 0.03 km s$^{-1}$, $V$=--21.6 $\pm$ 0.08 km s$^{-1}$, and 
$W$=--5.8 $\pm$ 0.19 km s$^{-1}$.  This is 1.4 $\sigma$ (1.8 $\pm$ 1.3 km s$^{-1}$) 
from the locus of Columba members from \citet{Torres:2008vq}.
Given the good kinematic agreement with Columba and appropriate lithium strength for the age of this group,
we adopt the previous membership assessment in Columba over BANYAN's field hypothesis. 

\textbf{2MASS J02490228--1029220} --- 
\citet{Bergfors:2015vs} identify lithium in this resolved triple system (\citealt{Janson:2012dc})  and find that its kinematics are a good match
to $\beta$ Pic, but the best hypothesis from our BANYAN-$\Sigma$ analysis is the field.
We detect lithium from our SOAR spectrum with a  strength of $\approx$310 m\AA, comparable to what Bergfors et al. measure. 
RVs for this system are presented in \citet{Durkan:2018dy} and support candidacy in $\beta$ Pic, although a parallax is needed 
to unambiguously confirm membership.   
We adopt previous assessments of this system as a candidate in $\beta$ Pic over BANYAN's field hypothesis

\textbf{2MASS J03520223+2439479} --- 
This star is a known member of the Pleiades (e.g., \citealt{Stauffer:2007gf}).
It has also been proposed as a member of Taurus, but \citet{Kraus:2017bg} show that its proper motion is
inconsistent with that region.
\citet{Walter:1988aa} and \citet{Soderblom:1993} measure lithium equivalent widths
of 350 and 302 m\AA, respectively.
The 0$\farcs$45 binary companion we uncovered with Robo-AO was first reported by \citet{Leinert:1993aa}.
$Gaia$ DR2 reports a parallax of 2.2 $\pm$ 0.7 mas ($\approx$450 pc), but the astrometric excess
noise parameter is  large (2.5 mas), implying the five-parameter astrometric solution is 
not an especially good fit to the data.  This is likely caused by acceleration from the binary companion
so the reported parallax is probably unreliable.   
We adopt previous assessments of this system as a member of the Pleiades over BANYAN's field hypothesis.

\textbf{2MASS J04435686+3723033} ---
 \citet{Schlieder:2010gka} identify this object and its wide ($\approx$9$''$) comoving companion 
 as likely members of the $\beta$ Pic moving group based on their activity and proper motions from SUPERBLINK.  
$\beta$ Pic membership is reaffirmed in \citet{Malo:2014bw}, \citet{Messina:2017cx}, and \citet{Shkolnik:2017ex}, 
but the best hypothesis from our BANYAN-$\Sigma$ analysis is the field.
Together with the \emph{Gaia} distance of 71.65 $\pm$ 0.26 pc and RV of --6.4 $\pm$ 0.2 from \citet{Malo:2014bw},
these proper motions imply $UVW$ space velocities of --10.66 $\pm$ 0.19 km s$^{-1}$, --19.08 $\pm$ 0.09 km s$^{-1}$, 
--8.40 $\pm$ 0.05 km s$^{-1}$.
These differ by 3.7 $\sigma$ (3.3 km $\pm$ 0.9 km s$^{-1}$) from the locus of $\beta$ Pic from \citet{Torres:2008vq}.
The M2 host star shows
modest lithium absorption (194 $\pm$ 4 m\AA \  from \citealt{Malo:2014bw} and $\approx$120 m\AA \ from our Mayall spectrum),
consistent with an age older than TWA but younger than Tuc-Hor.  
We adopt previous assessments of this system as a candidate member of $\beta$ Pic over BANYAN's field hypothesis.

\textbf{2MASS J05363633+2139330} ---
\citet{Li:1998bk} first identified this star as a candidate member of Taurus from its activity and
strong lithium absorption (480 m\AA).  
We also detect deep lithium in this star with an EW of $\approx$460 m\AA, but found a spectral type of
M2 which differs from the K4 classification by \citet{Li:1998bk}.
\citet{Mamajek:2016xx} suggest this star is a member of the proposed subgroup 118 Tau, which is
also suggested as the best hypothesis from BANYAN-$\Sigma$.
Membership in the broader Taurus complex was recently confirmed with a detailed analysis by \citet{Kraus:2017bg};
they also find this subgroup may be kinematically related to Taurus.
The proper motion and distance for this star from $Gaia$ DR2 is 
$\mu_{\alpha}$cos$\delta$=10.65 $\pm$ 0.19 mas yr$^{-1}$, 
$\mu_{\delta}$=--41.23 $\pm$ 0.14 mas yr$^{-1}$, 108.22 $\pm$ 1.59 pc, respectively,
similar to the other 118 Tau group members from \citet{Mamajek:2016xx}
($\mu_{\alpha}$cos$\delta$ $\approx$ +4 mas yr$^{-1}$;
$\mu_{\delta}$ $\approx$ --39 mas yr$^{-1}$; $d$ $\approx$ 120 pc).
 Given this consistent sky position, proper motion, and distance, we adopt  
 candidacy in 118 Tau as suggested by \citet{Mamajek:2016xx} and BANYAN.

 \begin{longrotatetable}
\startlongtable
\begin{deluxetable*}{lcccccccc}
\renewcommand\arraystretch{0.9}
\tabletypesize{\footnotesize}
\setlength{ \tabcolsep }{.1cm} 
\tablewidth{0pt}
\tablecolumns{9}
\tablecaption{Properties of Robo-AO Visual Binaries\label{tab:vbymg}}
\tablehead{
       \colhead{2MASS} & \colhead{$\mu_{\alpha} \cos(\delta)$\tablenotemark{a}}  & \colhead{$\mu_{\delta}$\tablenotemark{a}}   & \colhead{Distance\tablenotemark{a}} & \colhead{Cand.} &  \colhead{Proj. Sep.} & \colhead{BANYAN $\Sigma$} & \colhead{Literature} & \colhead{YMG} \\
       \colhead{Name} & \colhead{(mas yr$^{-1}$)}    & \colhead{(mas yr$^{-1}$)}  & \colhead{(pc)}  & \colhead{Comp.} & \colhead{('')}    & \colhead{Best Hyp.} & \colhead{YMG} & \colhead{Ref.} 
  }
\startdata
 J00074264+6022543    &     321.81   $\pm$   1.48                  &     --5.61    $\pm$   1.18                   &      16.48  $\pm$       0.25 &   CC1           &   0.86           &    Field        &    $\cdots$  & $\cdots$ \\
 J00133841+5245050    &      61.36   $\pm$   0.04                  &    --42.18    $\pm$   0.03                   &      54.63  $\pm$       0.09 &   CC1/CC2       &   3.13/0.86      &    Field        &    $\cdots$  & $\cdots$ \\
 J00160486+2319090    &     140.26   $\pm$   0.14                  &    --45.65    $\pm$   0.17                   &      37.88  $\pm$       0.14 &   CC1           &   2.66           &    Field        &    $\cdots$  & $\cdots$ \\
 J00164045+3000598    &     223.82   $\pm$   0.29                  &      23.70    $\pm$   0.21                   &      41.56  $\pm$       0.30 &   CC1           &   0.98           &    Field        &    $\cdots$  & $\cdots$ \\
 J00165678+2003551    &     212.24   $\pm$   0.44                  &    --28.39    $\pm$   0.58                   &      39.06  $\pm$       0.38 &   CC1           &   1.03           &    Field        &    $\cdots$  & $\cdots$ \\
\enddata
\end{deluxetable*}
\end{longrotatetable}

\textbf{2MASS J05374649+0231264} ---
\citet{daSilva:2009eu} first identify this lithium-rich ($EW$=300 m\AA) star as a member of Columba, which was
bolstered by \citet{Elliott:2016dd}.
We measure a somewhat lower lithium strength of $\approx$190~m\AA \ from our low-resolution Mayall spectrum.
Using the proper motion, distance (68.44 $\pm$ 0.19 pc), and RV (20.8 $\pm$ 2.8 km s$^{-1}$) from $Gaia$ DR2, 
the space velocities of this system are $U$=--13.1 $\pm$ 2.5 km s$^{-1}$, $V$=--20.6 $\pm$ 1.0 km s$^{-1}$, and 
$W$=--6.5 $\pm$ 0.7 km s$^{-1}$.  This is 1.0 $\sigma$ (1.3 $\pm$ 1.3 km s$^{-1}$) 
from the locus of Columba members from \citet{Torres:2008vq}.
Given the good kinematic agreement with Columba and appropriate lithium strength for the age of this group,
we adopt previous membership assessment in Columba over BANYAN's field hypothesis. 

\textbf{2MASS J05500858+0511536} ---
We measure modest lithium ($\approx$120 m\AA) in this little-studied active M2 star.
The best hypothesis from BANYAN-$\Sigma$ is Columba.
Using the proper motion, distance (64.45 $\pm$ 0.17 pc), and RV (18 $\pm$ 4 km s$^{-1}$) from $Gaia$ DR2, 
the space velocities of this system are $U$=--11.2 $\pm$ 3.7 km s$^{-1}$, $V$=--19.2 $\pm$ 1.4 km s$^{-1}$, and 
$W$=--5.2 $\pm$ 0.8 km s$^{-1}$.  This is 1.3 $\sigma$ (3.3 $\pm$ 2.7 km s$^{-1}$) 
from the locus of Columba members from \citet{Torres:2008vq}.
The absolute $V$-band magnitude of 8.6 mag and $V$--$J$ color of 3.3 mag place this star above the 
main sequence, in good agreement with other Columba members from \citet{Bell:2015gw}.
Overall this star appears to be an excellent new candidate member of Columba, but  
a more precise RV and lithium equivalent width measurement is needed for confirmation.

\textbf{2MASS J09595765--7221472} --
\citet{Elliott:2014is} identify this lithium-rich  star as a K4 candidate member of Carina, but 
the best hypothesis from BANYAN-$\Sigma$ is the field.
We find a somewhat later spectral type of K7 from our SOAR spectra. 
Our lithium measurement ($EW$ $\approx$ 270 m\AA) is comparable to that of \citet{Torres:2006bw} ($EW$ = 330 m\AA)
and suggests an age between TWA and AB Dor (e.g., \citealt{Murphy:2018bq}).
The velocities of this system are $U$=--8.8 $\pm$ 0.07 km s$^{-1}$, $V$=----21.6 $\pm$ 0.18 km s$^{-1}$, and 
$W$=--2.1 $\pm$ 0.05 km s$^{-1}$.  This is 2.5 $\sigma$ (3.0 $\pm$ 1.2 km s$^{-1}$) 
from the locus of Carina members from \citet{Torres:2008vq}.
This star is a better match to Tuc-Hor in terms of space motion, but is several tens of parsecs from 
established members of that group.
We adopt previous assessments of this system as a member of Carina over BANYAN's field hypothesis.

\textbf{2MASS J10260210--4105537}
This lithium-rich early M dwarf was proposed as a member of TWA by \citet{Bell:2015gw}, \citet{Pecaut:2013ej}, and \citet{Naud:2017fk}.
\citet{Gagne:2017gy} suggest it is a likely contaminant from Lower Centaurus Crux (LCC).
Using the new $Gaia$ DR2 distance of 84.9 $\pm$ 2 pc, the best hypothesis from BANYAN-$\Sigma$ is the field.
We measure a spectral type of M2 and strong lithium ($EW$ $\approx$ 410 m\AA), which is slightly less than that found by 
\citet[$EW$ = 500 $\pm$ 70 m\AA]{Rodriguez:2011gb}.
We also find unusually strong H$\alpha$ (EW$\approx$--10.4 \AA) above the 
envelope of saturated chromospheric emission identified by \citet{BarradoyNavascus:2003vz},
suggesting it may originate from ongoing accretion.
The distance and sky position of this object is more consistent with TWA than LCC (e.g., \citealt{Murphy:2015kv}) so we 
adopt previous assessments of this system as a likely member of TWA  over LCC and BANYAN's field hypothesis.
However, a radial velocity is needed to unambiguously establish group membership.

\textbf{2MASS J12003688--6337055} ---
This active, lithium-rich (EW$\approx$480 m\AA) M0 star was 
flagged as a likely LCC member using BANYAN-$\Sigma$.
Using the proper motion, distance (101.2 $\pm$ 0.3 pc), and RV (14.3 $\pm$ 1.8 km s$^{-1}$) from $Gaia$ DR2, 
the space velocities of this system are $U$=--9.7 $\pm$ 0.8 km s$^{-1}$, $V$=--21.06 $\pm$ 1.6 km s$^{-1}$, and 
$W$=--8.02 $\pm$ 0.06 km s$^{-1}$.  This is 1.1 $\sigma$ (2.7 $\pm$ 2.5 km s$^{-1}$) 
from the locus of LCC members from \citet{Gagne:2018wf}.
The $V$-band absolute magnitude of this star is 7.0 mag, which is about a magnitude above the main
sequence at the $V$--$J$ color of this object (2.7 mag; \citealt{Bell:2015gw}).
The sky position, space motion, lithium strength, and overluminosity are in excellent 
agreement with LCC.

\textbf{2MASS J12281909--7306346} ---
This active M0 star has  strong lithium absorption ---  $EW$$\approx$440 m\AA \ from our low-resolution Goodman spectrum --- 
and  has a best hypothesis of $\epsilon$ Cha from BANYAN-$\Sigma$.
It does not appear to be a previously known young star.
The space velocities of this system from $Gaia$ DR2 astrometry are $U$=--8.5 $\pm$ 0.7 km s$^{-1}$, $V$=--21.0 $\pm$ 1.0 km s$^{-1}$, and 
$W$=--7.9 $\pm$ 0.3 km s$^{-1}$.  This is 2.4 $\sigma$ (3.7 $\pm$ 1.5 km s$^{-1}$) 
from the locus of $\epsilon$ Cha members from \citet{Gagne:2018wf}.
We also note that this star's kinematics and distance (107 $\pm$ 2 pc) line up well  with the locus of LCC members 
(0.9 $\sigma$, or 1.9 $\pm$ 2.2 km s$^{-1}$).  Its sky position is just beyond the canonical (albeit arbitrarily defined) southern boundary of LCC at $b$=--10,
but all other indicators agree well with that association.  
We conclude that this star could plausibly belong to $\epsilon$ Cha or LCC, though the extended LCC is a better kinematic match.

\textbf{2MASS J12445897--6026409} ---
This M1 star was identified as a potential member of LCC using BANYAN-$\Sigma$.
We measure strong lithium absorption ($EW$$\approx$340 m\AA) consistent with LCC members of 
this spectral type.
The space velocities of this system from $Gaia$ DR2 astrometry are $U$=--6.7 $\pm$ 0.4 km s$^{-1}$, $V$=--18.3 $\pm$ 0.3 km s$^{-1}$, and 
$W$=--4.5 $\pm$ 0.3 km s$^{-1}$.  This is 1.1 $\sigma$ (3.8 $\pm$ 3.4 km s$^{-1}$) 
from the locus of LCC members from \citet{Gagne:2018wf}.
We conclude that this star is a previously unrecognized member LCC.

\textbf{2MASS J13314666+2916368} --- 
We measure unusually strong H$\alpha$ emission of --16.2 \AA \ from this M5 star, suggesting it may
originate from ongoing accretion.  The parallactic distance from $Gaia$ is 18.3 pc.
\citet{Riedel:2014ce} identify this close binary as a possible member of Carina or Columba.
If it is a member of either of these groups and if the strong H$\alpha$ originates from ongoing
accretion, this would be an unusually long disk dissipation timescale possibly similar to
the peculiar system found by \citet{Murphy:2018bq}.

\textbf{2MASS J13493313--6818291} ---
\citet{Malo:2013gn} identify this active M dwarf as a candidate member of Argus, but we find that the best hypothesis
from BANYAN-$\Sigma$ is LCC.
\citet{Janson:2012dc} resolve it into a close visual triple.
We measure a spectral type of M3 and find strong lithium ($EW$$\approx$360 m\AA) from our moderate-resolution Goodman spectrum,
implying an age significantly younger than Argus ($\approx$40--50~Myr; \citealt{Zuckerman:2018aa}).
The distance (99.8 $\pm$ 1.5 pc) and proper motion ($\mu_{\alpha}$cos$\delta$=--31.1 $\pm$ 0.2 mas yr$^{-1}$,  
$\mu_{\delta}$=--19.7 $\pm$ 0.2 mas yr$^{-1}$) is in good agreement
with LCC.  We conclude that this star is most likely an LCC member, but an RV is needed
for confirmation.  

\textbf{2MASS J15354856--2958551} ---
This M4 star is noteworthy for having the strongest H$\alpha$ emission (EW$\approx$--43 \AA) of any
star for which we obtained a spectrum in this program, indicating active disk accretion.  
\citet{Brandner:1996vc} resolve this star into a 0$\farcs$9 binary and 
\citet{Barenfeld:2016cc} detect the disk in continuum and CO line emission with ALMA.
\citet{Koeler:2000aa} identify this star as a member of USco,
but our BANYAN-$\Sigma$ analysis suggests it is a field star based on the UCAC4 proper motion
(no astrometric solution is presented in $Gaia$ DR2).
We measure strong lithium with an $EW$ of $\approx$500 m\AA, implying a
young age consistent with members of the Sco-Cen complex and certainly less than a few tens of 10~Myr.
We note that the sky position and proper motion align with UCL.
We conclude that this star is a good candidate for UCL, but an RV and parallax are needed 
to fully assess membership in this subgroup.  

\textbf{2MASS J15451903--4431361} ---
This little-studied active M3 star shows strong lithium absorption ($EW$$\approx$380 m\AA) and was 
identified as a candidate UCL member using BANYAN-$\Sigma$.
The sky position and proper motion are in good agreement with UCL membrs, but the 
distance from $Gaia$ DR2 of 89.3 $\pm$ 3.7 pc is much closer the vast majority of established
members (\citealt{Wright:2018aa}).  However, this does not exclude candidacy in that subgroup
because our targets are intentionally biased to closer distances which would naturally sample the
closest members of this complex.
We also note that the $Gaia$ DR2 excess noise parameter for this target is quite large
(3.1 mas), which may point to an unseen companion that could be affecting the five-parameter astrometric fit.
We conclude that this star may be an unusually nearby member of UCL, but an RV (and perhaps 
better parallax solution) is needed for confirmation.

\textbf{2MASS J16430128--1754274} --- 
This active M1 star has been widely listed as a kinematic member of $\beta$ Pic  (e.g., \citealt{Kiss:2010cb}; \citealt{Binks:2014gd}; \citealt{Shkolnik:2017ex}).
However, the best hypothesis from BANYAN-$\Sigma$ is the field and 
it received a low membership probability in $\beta$ Pic by \citet{Malo:2013gn}.
This star has strong lithium absorption, with $EW$ measurements of 300 $\pm$ 20 m\AA \ by \citet{Kiss:2010cb},
364 $\pm$ 20 m\AA \ by \citet{Binks:2014gd}, and $\approx$280 m\AA \ in this work from our low-resolution RC-Spec spectrum.
Based on the $Gaia$ DR2 distance of 71.1 $\pm$ 0.3 pc and RV of --9.3 $\pm$ 0.4 km s$^{-1}$ from \citet{Malo:2014dk}, 
the space velocities of this system are $U$=--7.6 $\pm$ 0.4 km s$^{-1}$, $V$=--20.1 $\pm$ 0.08 km s$^{-1}$, and 
$W$=--5.7 $\pm$ 0.13 km s$^{-1}$.  This is 5.0 $\sigma$ (6.0 $\pm$ 1.2 km s$^{-1}$) 
from the locus of $\beta$ Pic members from \citet{Torres:2008vq}.
We conclude that this star is a poor match with $\beta$ Pic and does not agree especially well with any other known nearby moving groups.
 
\textbf{2MASS J16455062+0343014} ---
\citet{Schlieder:2012gj} and \citet{Schlieder:2012gu} identify this active M dwarf as a likely member of AB Dor, but
 the best hypothesis from BANYAN-$\Sigma$ is the field.
 We measure a spectral type of M2 and find modest lithium absorption ($EW$ $\approx$ 120 m\AA).
We also resolve this source into a 2$''$ binary with Robo-AO and confirm that the pair are physically bound.
Based on the $Gaia$ DR2 distance of 44.89 $\pm$ 0.08 pc and RV of --15.5 $\pm$ 0.7 km s$^{-1}$ from \citet{Schlieder:2012gu}, 
the space velocities of this system are $U$=--2.3 $\pm$ 0.6 km s$^{-1}$, $V$=--26.3 $\pm$ 0.2 km s$^{-1}$, and 
$W$=--11.2 $\pm$ 0.3 km s$^{-1}$.  This is 3.4 $\sigma$ (5.0 $\pm$ 1.5 km s$^{-1}$) 
from the locus of AB Dor members from \citet{Torres:2008vq}.
However, when we use the RV of --21.7 $\pm$ 1.8 km s$^{-1}$ from $Gaia$ DR2, the
space velocities of this system are $U$=--7.4 $\pm$ 1.4 km s$^{-1}$, $V$=--28.2 $\pm$ 0.6 km s$^{-1}$, and 
$W$=--14.2 $\pm$ 0.9 km s$^{-1}$, or only 0.9 $\sigma$ (1.5 $\pm$ 1.6 km s$^{-1}$) 
from the locus of AB Dor members.  
We conclude that this visual and spectroscopic binary remains an excellent candidate member of AB Dor.  Longer-baseline
RV monitoring will be useful to measure a systemic velocity for this pair.
 
\textbf{2MASS J17213497--2152283} ---
We measure strong H$\alpha$ emission (EW$\approx$--13.4 \AA) and lithium absorption (EW$\approx$370 m\AA)
in this little-studied active M4 star.  
The best hypothesis from BANYAN-$\Sigma$ is UCL, but the sky position lies at the eastern edge of USco 
and disagrees with the UCL subgroup.
However, the distance from $Gaia$ DR2 of 101.0 $\pm$ 0.7 pc places it closer than nearly all USco members (\citealt{Wright:2018aa}).
We conclude that this star is likely related to the Sco-Cen complex, but perhaps not directly associated
with the canonically defined subgroups. 

\textbf{2MASS J23093711--0225551} ---
This active K4 star was identified as a candidate member of Carina by \citet{Elliott:2014is}
but the best hypothesis from BANYAN-$\Sigma$ is the field.  
Based on parallactic distance of 52.6 $\pm$ 0.4 pc and RV of --12.7 $\pm$ 0.4 km s$^{-1}$ from $Gaia$ DR2, 
the space velocities of this star are $U$=--9.64 $\pm$ 0.09 km s$^{-1}$, $V$=--20.8 $\pm$ 0.2 km s$^{-1}$, and 
$W$=--0.2 $\pm$ 0.3 km s$^{-1}$.  This is 3.4 $\sigma$ (4.7 $\pm$ 1.4 km s$^{-1}$) 
from the locus of Carina members from \citet{Torres:2008vq}.
The kinematics are in good agreement with Tuc-Hor, but this star would be a spatial outlier if it belongs to that group.
We measure weak lithium (EW $\approx$ 130 m\AA) from our low-resolution Goodman spectrum, implying an age older than $\beta$ Pic but
consistent with scatter in Tuc-Hor and AB Dor.  
We conclude that this star is most consistent with the field, but could be a kinematic outlier of Carina or perhaps a spatial outlier of Tuc-Hor.

\section{Summary}{\label{sec:discussion}

The goal of this study is to identify new young stars in the solar neighborhood for future direct imaging surveys
of exoplanets.  
We began with a sample of 2060
late-K through early-M dwarfs selected on the basis of X-ray and UV activity cuts,
proper motions, NIR color cuts, and optical brightness.
Follow-up low-resolution optical spectra were obtained for 632 stars, 58 of which show strong lithium 
absorption.  
Among the lithium-rich stars, 34 are previously known members of nearby moving groups while seven are new.
The rest appear to be young field stars without any obvious connection to an established kinematic group.
We also acquired Robo-AO observations of 1011 northern stars in our sample of active K/M dwarfs;
239 of these have nearby point sources within 4$''$, the majority of which are likely to be physical companions.
Many of these have kinematics consistent with young moving groups which long-baseline RV monitoring
can better constrain by measuring systemic RVs.

\acknowledgments

It is a pleasure to thank Diane Harmer, Sean Points, and all support staff and telescope operators at KPNO and CTIO who helped make these observations possible.
CZ is supported by a Dunlap Fellowship at the Dunlap Institute for Astronomy \& Astrophysics, funded through an endowment established by the Dunlap family and the University of Toronto. VS was supported part from the 
John W. Cox Endowment for the Advanced Studies in Astronomy.
This publication makes use of data products from the Two Micron All Sky Survey, which is a joint project of the University of Massachusetts and the Infrared Processing and Analysis Center/California Institute of Technology, funded by the National Aeronautics and Space Administration and the National Science Foundation
This work has made use of data from the European Space Agency (ESA) mission
{\it Gaia} (\url{https://www.cosmos.esa.int/gaia}), processed by the {\it Gaia}
Data Processing and Analysis Consortium (DPAC,
\url{https://www.cosmos.esa.int/web/gaia/dpac/consortium}). Funding for the DPAC
has been provided by national institutions, in particular the institutions
participating in the {\it Gaia} Multilateral Agreement.
This publication makes use of data products from the Wide-field Infrared Survey Explorer, which is a joint project of the University of California, Los Angeles, and the Jet Propulsion Laboratory/California Institute of Technology, funded by the National Aeronautics and Space Administration.
The Robo-AO system was developed by collaborating partner institutions, the California Institute of Technology 
and the Inter-University Centre for Astronomy and Astrophysics, and with the support of the National Science Foundation 
under Grant Nos. AST-0906060, AST-0960343 and AST-1207891, the Mt. Cuba Astronomical Foundation and by a gift 
from Samuel Oschin. Ongoing science operation support of Robo-AO is provided by the California Institute of Technology 
and the University of Hawai`i. C.B. acknowledges support from the Alfred P. Sloan Foundation.
 NASA's Astrophysics Data System Bibliographic Services together with the VizieR catalogue access tool and SIMBAD database 
operated at CDS, Strasbourg, France, were invaluable resources for this work.
Based on observations made with the NASA Galaxy Evolution Explorer. 
GALEX is operated for NASA by the California Institute of Technology under NASA contract NAS5-98034.
This research has made use of the Washington Double Star Catalog maintained at the U.S. Naval Observatory.

Based in part on observations obtained at the Southern Astrophysical Research (SOAR) telescope
(NOAO Prop. ID 2013B-0496, 2014A-0019, 2015A-0016; PI: B. Bowler), 
which is a joint project of the Minist\'{e}rio da Ci\^{e}ncia, Tecnologia, Inova\c{c}\~{a}os e Comunica\c{c}\~{a}oes (MCTIC) do Brasil, 
the U.S. National Optical Astronomy Observatory (NOAO), the University of North Carolina at Chapel Hill (UNC), 
and Michigan State University (MSU).
Based in part on observations at Kitt Peak National Observatory, National Optical Astronomy Observatory 
(NOAO Prop. ID 2013B-0496, 2014A-0019, 2015A-0016; PI: B. Bowler), 
which is operated by the Association of Universities for Research in Astronomy (AURA) 
under cooperative agreement with the National Science Foundation. 
Institutional allocation for the Robo-AO observations at the P60 telescope was provided based on the prior affiliation 
of BPB and SH with the California Institute of Technology.
The authors are honored to be permitted to conduct astronomical research on Iolkam Du'ag (Kitt Peak), 
a mountain with particular significance to the Tohono O'odham.

\facilities{Mayall (RC-Spec), SOAR (Goodman Spectrograph), PO:1.5m (Robo-AO), UH:2.2m (SNIFS)}

\newpage


\clearpage

\newpage

\end{document}